\documentclass[aps,prd,letterpaper,showkeys,showpacs,twocolumn,nofootinbib,preprintnumbers,longbibliography,unsortedaddress]{revtex4-1}

\usepackage[top=2.8cm, bottom=2.8cm, left=2.4cm, right=2.4cm]{geometry}
\usepackage[utf8]{inputenc}  
\usepackage{amsmath,amssymb,lmodern,amsfonts} 
\usepackage{graphicx}
\usepackage{hyperref}
\usepackage{color}
\usepackage{natbib}
\usepackage[caption=false]{subfig}

\newcommand{\lsim}{\lesssim}
\newcommand{\gsim}{\gtrsim}

\begin{document}

\preprint{PI/UAN-2019-645FT}

\title{Einstein Yang-Mills Higgs dark energy revisited}

\author{Miguel Álvarez}
\email{miguel.alvarez1@correo.uis.edu.co}
\affiliation{Escuela  de  F\'isica,  Universidad  Industrial  de  Santander, \\ Ciudad  Universitaria,  Bucaramanga  680002,  Colombia}

\author{J. Bayron Orjuela-Quintana}
\email{john.orjuela@correounivalle.edu.co}
\affiliation{Departamento  de  F\'isica,  Universidad  del Valle, \\ Ciudad  Universitaria Meléndez,  Santiago de Cali  760032,  Colombia}

\author{Yeinzon Rodr\'iguez}
\email{yeinzon.rodriguez@uan.edu.co}
\affiliation{Centro de Investigaciones en Ciencias B\'asicas y Aplicadas, Universidad Antonio Nari\~no, \\ Cra 3 Este \# 47A-15, Bogot\'a D.C. 110231, Colombia}
\affiliation{Escuela  de  F\'isica,  Universidad  Industrial  de  Santander, \\ Ciudad  Universitaria,  Bucaramanga  680002,  Colombia}
\affiliation{Simons Associate at The Abdus Salam International Centre for Theoretical Physics, \\ Strada Costiera 11, I-34151, Trieste, Italy}

\author{César A. Valenzuela-Toledo}
\email{cesar.valenzuela@correounivalle.edu.co}
\affiliation{Departamento  de  F\'isica,  Universidad  del Valle, \\ Ciudad  Universitaria Meléndez,  Santiago de Cali  760032,  Colombia}


\begin{abstract}
Inspired in the Standard Model of Elementary Particles, the Einstein Yang-Mills Higgs action with the Higgs field in the SU(2) representation was proposed in {\it Class. Quantum Grav.} {\bf 32} (2015) 045002  as the element responsible for 
the dark energy phenomenon. 
We revisit this action emphasizing in a very important aspect not sufficiently explored in the original work and that substantially changes its conclusions.  This aspect is the role that the Yang-Mills Higgs interaction plays at fixing the gauge for the Higgs field, in order to sustain a homogeneous and isotropic background, and at driving the late accelerated expansion of the Universe by moving the Higgs field away of the minimum of its potential and holding it towards an asymptotic finite value.  We analyse the dynamical behaviour of this system and supplement this analysis with a numerical solution whose initial conditions are in agreement with the current observed values for the density parameters.  This scenario represents a step towards a successful merging of cosmology and well-tested particle physics phenomenology.
\end{abstract}

\pacs{98.80.Cq; 95.36.+x}

\keywords{dark energy; non-Abelian gauge field theories; Higgs field}

\maketitle

\section{Introduction}

The nature of the late accelerated expansion of the Universe is one of the most important, difficult, and fundamental problems in Physics.  Discovered in 1998 by a couple of supernova projects \cite{Riess:1998cb,Perlmutter:1998np}, the present accelerated expansion of the Universe resurrected the Einstein's {\it ``biggest blunder''} and has consolidated \cite{Huterer:2017buf} as a key element in the best cosmological description of our Universe:  the $\Lambda$ Cold Dark Matter ($\Lambda$CDM) model (also known as the Standard Cosmological Model) \cite{Amendola:2015ksp,Aghanim:2018eyx}.  Although $\Lambda$, the cosmological constant, is a very simple and efficient element that can drive the present accelerated expansion of the Universe, it looked ugly and incomprehensible to Einstein and, indeed, any attempt to describe its nature as the vacuum energy density in a particle physics context has badly failed \cite{Amendola:2015ksp,Weinberg:1988cp}.  The profuse search for alternatives to the cosmological constant has basically split into two categories:  the introduction of non-standard forms of matter and modifications to the Einstein gravity.  The former implies some violation of the energy conditions while the latter has been recently under observational pressure \cite{Collett:2018gpf,Ezquiaga:2018btd,He:2018oai}.  In contrast, little has been done in the search for a mechanism that mimics the cosmological constant while involving standard forms of matter that minimally couple to the gravity described by General Relativity.  Inspired in the Standard Model of Particle Physics (SM) (see, for example, Refs. \cite{Weinberg:1996kr,Kane:1987gb}), the Einstein Yang-Mills Higgs action was introduced in Refs. \cite{Moniz:1991kx,Ochs:1996yr} in order to study its role in the description of the primordial inflationary period.  Many years later, M. Rinaldi studied such an action in Ref. \cite{Rinaldi:2014yta} arguing that a late period of accelerated expansion was possible due to the complex doublet nature of the Higgs field, charged under the SU(2) gauge symmetry\footnote{It is worth stressing that this Higgs is not the one of the SM whose discovery in 2012 was reported in Refs. \cite{Aad:2012tfa,Chatrchyan:2012xdj}.}. In summary, Rinaldi believed to have shown that, when considering the effect of gravity, the Higgs complex field rotates about the axis of the Mexican hat potential, away from the minimum of its potential but slowly chasing it, giving way to an slowly-varying effective cosmological constant much in the same way as in the {\it ``spintessence''} scenario \cite{Boyle:2001du}.  Despite the fact that the Yang-Mills fields, charged under SU(2), were considered, Rinaldi neglected the interaction between them and the Higgs field, arriving to the conclusion that the former behaved as a pure radiation fluid.  Later on, in Ref. \cite{Rinaldi:2015iza}, the same author took into account such an interaction, but this time in the SO(3) representation for the Higgs field, arriving to similar conclusions as in Ref. \cite{Rinaldi:2014yta}\footnote{A careful analysis of Ref. \cite{Rinaldi:2015iza} shows that its assumption of a homogeneous and isotropic background is incorrect, the latter being actually described by an axisymmetric Bianchi I metric at best \cite{newpaper} (M. Álvarez, Y. Rodríguez and M. Rinaldi, private communications).}.  Our purpose in this paper is to seriously consider this interaction in the SU(2) representation for the Higgs field and determine the extent to which it affects the dark energy mechanism.  The layout of this paper is the following:  in Section \ref{motivation}, we will present the Einstein Yang-Mill Higgs action with the Higgs complex field in the SU(2) representation and show that the interaction between the Yang-Mills fields and the Higgs field cannot be neglected since it has very important effects that substantially change Rinaldi's conclusions;  in Section \ref{equations}, the respective equations of motion will be derived;  in Section \ref{dynsys}, a dynamical system analysis will be performed with the purpose of obtaining the main qualitative features of the time evolution in this scenario;  this analysis will be complemented in Section \ref{numerical} with a numerical solution whose initial conditions are in agreement with the present observed values of the density parameters; just before finishing, some remarks about observational signatures as well as implications for the particle physics phenomenology will be discussed in Sections \ref{signatures} and \ref{higgsmass} respectively;  finally, our conclusions will be presented in Section \ref{conclusions}.

\section{Importance of the Yang-Mills Higgs interaction} \label{motivation}

The Einstein Yang-Mills Higgs action \cite{Moniz:1991kx,Ochs:1996yr,Rinaldi:2014yta} borrows some elements from the SM (see, for example, Refs. \cite{Weinberg:1996kr,Kane:1987gb}) such as the SU(2) gauge invariance -- and, therefore, the existence of three gauge bosons -- and the presence of a Higgs complex doublet.  It also involves non-interacting radiation and matter fluids, and the gravity is described by the General Relativity.  Its action is given by
\begin{eqnarray}
S = && \int d^4 x \ \sqrt{-g} \ \Big[ \frac{m_P^2}{2} R - \frac{1}{4} F^{\mu \nu}_a F_{\mu \nu}^a - (D^\mu \Phi)^\dagger (D_\mu \Phi) \nonumber \\
&& - V(\Phi^2) + \mathcal{L}_r + \mathcal{L}_m \Big] \,,
\end{eqnarray}
where $g$ is the determinant of the space-time metric, $m_P$ is the reduced Planck mass, $R$ is the Ricci scalar, $F_{\mu \nu}^a$ is the non-Abelian gauge field strength tensor defined by
\begin{equation}
F_{\mu \nu}^a \equiv \partial_\mu A_\nu^a - \partial_\nu A_\mu^a + \gamma \epsilon^a_{\,\,\, bc} A_\mu^b A_\nu^c \,,
\end{equation}
$A_\mu^a$ representing the vector fields, $\gamma$ being the SU(2) group coupling constant, and $\epsilon_{abc}$ being the three-dimensional Levi-Civita symbol, $D_\mu$ is the gauge covariant derivative defined by
\begin{equation}
D_\mu \equiv \nabla_\mu - i \gamma \frac{\sigma_c}{2} A_\mu^c \,,
\end{equation}
$\nabla_\mu$ being the space-time covariant derivative and $\sigma_c$ denoting the Pauli matrices, $\Phi$ is the Higgs complex doublet described as
\begin{equation}
\Phi \equiv
\begin{pmatrix}
\Phi_1 \\
\Phi_2
\end{pmatrix}
=
\begin{pmatrix}
\phi_1 + i \psi_1 \\
\phi_2 + i \psi_2 
\end{pmatrix} \,,
\end{equation}
$\phi_1,\phi_2,\psi_1$ and $\psi_2$ being real scalar fields, $V(\Phi^2)$ is the usual Higgs Mexican hat potential given by
\begin{equation}
V(\Phi^2) \equiv \frac{\lambda}{4} (\Phi^2 - \Phi_0^2)^2 \,,
\end{equation}
$\lambda$ being the quartic coupling constant and $\Phi_0$ denoting the Higgs field configuration at the minimum of its potential, and $\mathcal{L}_r, \mathcal{L}_m$ are the Lagrangian densities for the radiation and matter fluids respectively.  Greek indices run from 0 to 3 and denote space-time components, and Latin indices run from 1 to 3 and denote SU(2) gauge components.

As is well known, our Universe is highly homogeneous and isotropic \cite{Aghanim:2018eyx}, which allows us to describe it at the background level by the Friedmann-Lemaitre-Robertson-Walker (FLRW) metric which in Cartesian coordinates reads as
\begin{equation}
ds^2 = -dt^2 + a^2(t) d\vec{x}^2 \,,
\end{equation}
where $a(t)$ is the expansion parameter as a function of the cosmic time $t$.  Such a configuration requires a homogeneous arrangement of fields and an isotropic energy-momentum tensor. The latter is defined as
\begin{equation}
T_{\mu \nu} \equiv 2 \frac{\partial \mathcal{L}_{\rm mat}}{\partial g^{\mu \nu}} - g_{\mu\nu} \mathcal{L}_{\rm mat} \,, \label{emdef}
\end{equation}
where 
\begin{equation}
\mathcal{L}_{\rm mat} = - \frac{1}{4} F^{\mu \nu}_a F_{\mu \nu}^a - (D^\mu \Phi)^\dagger (D_\mu \Phi) - V(\Phi^2) + \mathcal{L}_r + \mathcal{L}_m \,,
\end{equation}
and $g_{\mu\nu}$ is the FLRW metric.

From the definition in Eq. (\ref{emdef}), it is obvious that the second term on the right member of the equation vanishes for $\mu\nu = \mu i$, with $\mu \neq i$, i.e., this term neither contributes to the momentum density nor to the anisotropic stress.  What it is not obvious is whether the first term on the right member of the equation contributes to these physical quantities.  A simple but delicate calculation shows that this first term is given by
\begin{equation}
2 \frac{\partial \mathcal{L}_{\rm mat}}{\partial g^{\mu \nu}} = - F_{\mu \rho}^a F_{\nu \;\; a}^{\,\,\, \rho} - 2 (D_{(\mu} \Phi)^\dagger(D_{\nu)} \Phi) + 2 \frac{\partial (\mathcal{L}_r + \mathcal{L}_m)}{\partial g^{\mu \nu}}\,,
\end{equation}
where the round brackets enclosing space-time indices mean a standard symmetrization:  $B_{(\mu} C_{\nu)} \equiv (B_\mu C_\nu + B_\nu C_\mu)/2$.  A close examination of the $ij$ components reveals that the only way to avoid anisotropic stress, assuming isotropic radiation and matter fluids, is by employing the {\it ``cosmic triad''} configuration\footnote{For other applications of the cosmic triad to dark energy, see Refs. \cite{ArmendarizPicon:2004pm,Mehrabi:2017xga,Rodriguez:2017wkg}.} for the gauge fields:
\begin{equation}
A_\mu^a = f \delta_\mu^a \,. \label{costriad}
\end{equation}
In fact,
\begin{equation}
2 \frac{\partial \mathcal{L}_{\rm mat}}{\partial g^{\mu \nu}} \Big|_{\mu\nu = ij} \supset \delta_{ij} \left( \dot{f}^2 - 2 \frac{\gamma^2 f^4}{a^2} - \frac{1}{2} \gamma^2 \Phi^2 f^2 \right) \,,
\end{equation}
when the cosmic triad is employed.
However,  we see that this is insufficient to avoid a contribution to the momentum density:
\begin{equation}
2 \frac{\partial \mathcal{L}_{\rm mat}}{\partial g^{\mu \nu}} \Big|_{\mu\nu = 0i} = -\frac{1}{4} \gamma f \operatorname{Im} (\dot{\Phi}^\dagger \sigma_i \Phi) \,.
\end{equation}
The origin of this contribution is, clearly, the interaction between the Yang-Mills fields and the Higgs field and, since it imposes severe restrictions on the Higgs doublet, {\em it cannot be neglected} as was done in Ref. \cite{Rinaldi:2014yta}.  Such restrictions are the following:
\begin{eqnarray}
\operatorname{Im} (\dot{\Phi}_1^* \Phi_2 + \dot{\Phi}_2^* \Phi_1) &=& 0 \,, \nonumber \\
\operatorname{Re} (\dot{\Phi}_1^* \Phi_2 - \dot{\Phi}_2^* \Phi_1) &=& 0 \,, \nonumber \\
\operatorname{Im} (\dot{\Phi}_1^* \Phi_1 - \dot{\Phi}_2^* \Phi_2) &=& 0 \,,
\end{eqnarray}
which are equivalent to
\begin{eqnarray}
\dot{\phi}_1 \psi_2 - \dot{\psi}_1 \phi_2 + \dot{\phi}_2 \psi_1 - \dot{\psi}_2 \phi_1 &=& 0 \,, \nonumber \\
\dot{\phi}_1 \phi_2 + \dot{\psi}_1 \psi_2 - \dot{\phi}_2 \phi_1 - \dot{\psi}_2 \psi_1 &=& 0 \,, \nonumber \\
\dot{\phi}_1 \psi_1 - \dot{\psi}_1 \phi_1 - \dot{\phi}_2 \psi_2 + \dot{\psi}_2 \phi_2 &=& 0 \,. \label{restrictions}
\end{eqnarray}
Thus, the additional condition to avoid an anisotropic energy-momentum tensor is to fix the gauge so that
 \begin{equation}
\Phi(t) \equiv
\begin{pmatrix}
\phi(t) \\
0
\end{pmatrix} \,, \label{gaufix}
\end{equation}
with $\phi$ being a real scalar field\footnote{Any gauge transformation of this configuration satisfies as well the restrictions in Eq. \ref{restrictions}.}.  This conclusion is clearly at odds with the claims in Ref. \cite{Rinaldi:2014yta} since any evolution of the Higgs field that involves a rotation around the axis of the Mexican hat potential, as in the spintessence model \cite{Boyle:2001du}, will immediately produce a huge amount of anisotropy in disagreement with observations.  

It is worth stressing that if it were not for the presence of the gauge fields and their interaction with the Higgs field, the energy-momentum tensor would be isotropic no matter the chosen gauge for the Higgs field.  In connection with this, we know that there are no preferred directions associated to scalar fields, so that cosmological models based solely on homogeneous versions of them always produce both background and statistical isotropy\footnote{Inhomogeneous scalar fields can drive a homogeneous but anisotropic universe if extra symmetries are added, as in the solid inflation model \cite{Endlich:2012pz,Bartolo:2013msa}.}.  In contrast, it is clear that vector fields enjoy inherent preferred directions, so that cosmological models based solely on them produce a huge amount of both background and statistical anisotropy unless the cosmic triad configuration is employed \cite{Rodriguez:2015xra};  this is the case of models like vector inflation \cite{Golovnev:2008cf}, and gauge-flation \cite{Maleknejad:2011jw,Nieto:2016gnp}.

Will this conclusion throw away the possibility of a dark energy mechanism for the Einstein Yang-Mills Higgs action?  The answer is no, as we will see in the following sections.

\section{Equations of motion} \label{equations}

Once the cosmic triad configuration has been employed, as in Eq. (\ref{costriad}), and the gauge for the Higgs field has been fixed, as in Eq. (\ref{gaufix}), the gravitational field equations in the FLRW background are the following:
\begin{eqnarray}
H^2 = \frac{1}{3m_P^2} \Big[ && \frac{3}{2} \frac{\dot{f}^2}{a^2} + \dot{\phi}^2 + \frac{3}{2} \frac{\gamma^2 f^4}{a^4} + \frac{3}{4} \frac{\gamma^2 \phi^2 f^2}{a^2} + V(\phi^2) \nonumber \\
&& + \rho_r + \rho_m \Big] \,, \label{fried} \\
\dot{H} = -\frac{1}{2m_P^2} \Big[ && 2\frac{\dot{f}^2}{a^2} + 2\dot{\phi}^2 + 2\frac{\gamma^2 f^4}{a^4} + \frac{\gamma^2 \phi^2 f^2}{2 a^2} \nonumber \\
&& + \frac{4}{3} \rho_r + \rho_m \Big] \,, \label{hdot}
\end{eqnarray} 
where $H \equiv \dot{a}/a$ is the Hubble parameter and $\rho_r,\rho_m$ are the energy densities of the radiation and matter fluids respectively.  These equations are supplemented by the field equations of motion obtained by varying the action with respect to $A_\mu^a$ and $\Phi$ and replacing the field configurations and the FLRW metric:
\begin{eqnarray}
\ddot{f} + H \dot{f} + 2 \frac{\gamma^2 f^3}{a^2} + \frac{\gamma^2 f \phi^2}{2} &=& 0 \,, \label{feq} \\
\ddot{\phi} + 3 H \dot{\phi} + \frac{3}{4} \frac{\gamma^2 f^2 \phi}{a^2} + \frac{d V(\phi)}{d \phi} &=& 0 \,. \label{phieq}
\end{eqnarray}
One of these four equations is redundant but, nevertheless, keeping the four equations will be useful for the construction and analysis of the autonomous dynamical system in the following section.  

An important aspect of Eq. (\ref{feq}) is that it reveals the existence of a potential for the gauge field, a condition that many physicists deny. This potential is clearly a product of the interaction between the gauge field and the Higgs field and cannot be neglected.  As we will see, $f=0$ minimizes this potential;  this is in agreement with the usual assumption in standard particle physics of assigning a vanishing vacuum expectation value to the gauge fields.  On the other hand, the same interaction between the gauge field and the Higgs field is the responsible for the existence of an effective potential for the Higgs field (see Eq. (\ref{phieq})) that produces an accelerated expansion period despite the fact the Mexican hat potential is not flat enough.  This kind of effective potentials is ubiquitous when dealing with multiple types of fields (see e.g. Ref. \cite{Dimopoulos:2010xq}).

\section{Dynamical system analysis} \label{dynsys}

With the purpose of analyzing the dynamical behaviour of the physical quantities in this scenario, we can construct an autonomous dynamical system by defining the following dimensionless dynamical quantities:
\begin{eqnarray}
x \equiv \frac{\dot{f}}{\sqrt{2} a m_P H} \,, && \hspace{5mm} y \equiv \frac{\gamma f^2}{\sqrt{2} a^2 m_P H} \,, \nonumber \\
w \equiv \frac{\gamma f \phi}{2 a m_P H} \,, && \hspace{5mm} z \equiv \frac{\dot{\phi}}{\sqrt{3} m_P H} \,, \nonumber \\
v \equiv \frac{1}{m_P H} \sqrt{\frac{V(\phi)}{3}} \,, && \hspace{5mm} r \equiv \frac{1}{m_P H} \sqrt{\frac{\rho_r}{3}} \,, \nonumber \\
m \equiv \frac{1}{m_P H} \sqrt{\frac{\rho_m}{3}} \,, && \hspace{5mm} l \equiv \frac{\sqrt{2} a m_P}{f} \,. \label{variables}
\end{eqnarray}
Thus, the Friedmann equation in Eq. (\ref{fried}) becomes the constraint equation
\begin{equation}
x^2 + y^2 + w^2 + z^2 + v^2 + r^2 + m^2 = 1 \,, \label{constraint}
\end{equation}
from which we can write the $m$ variable as a function of the other variables.

Exchanging the cosmic time by the number of e-folds $N$ defined by $dN = H dt$, and taking into account Eqs. (\ref{hdot}) - (\ref{constraint}), we can write the evolution equations for each independent dimensionless variable as follows:
\begin{eqnarray}
x' &=& x(q - 1) - l (2y^2 + w^2) \,, \label{xeq} \\
y' &=& y(2xl + q -1) \,, \\
w' &=& w(x l + q) + \frac{\sqrt{3}}{2} l y z \,, \\
z' &=& z(q - 2) - w l \left(2\alpha v + \frac{\sqrt{3}}{2}y \right) \,, \\
v' &=& v(q + 1) + \alpha l w z \,, \\
r' &=& r(q - 1) \,, \\
l' &=& l(1 - l x) \,, \label{leq}
\end{eqnarray}
where a prime means a derivative with respect to $N$, $q \equiv - \ddot{a} a/ \dot{a}^2$ is the deceleration parameter which is given by
\begin{equation}
q = \frac{1}{2} (1 + x^2 + y^2 - w^2 + 3z^2 - 3v^2 + r^2) \,, \label{q}
\end{equation}
and $\alpha$ is the positive dimensionless constant defined by $\alpha \equiv \sqrt{\lambda/2\gamma^2}$.  This autonomous dynamical system is similar to the one obtained by Rinaldi in Ref. \cite{Rinaldi:2015iza} where he studied the Einstein Yang-Mills Higgs action with the Higgs field in the SO(3) representation;  however, the conclusions obtained in this section differ in some important aspects from his.

The following list shows the critical manifolds and critical points of the system and gives a discussion about each of them (the complete list of the respective eigenvalues and eigenvectors is shown in the Appendix \ref{appeigen}).  But, before that, it is important to notice that, although the variables in Eq. (\ref{variables}) are the ones that describe the dynamical system, the actual quantities we are interested in are the physical ones, i.e., the physical vector field $f/a$, its speed $(f/a)'$, the Higgs field $\phi$, its speed $\phi'$, and the deceleration parameter $q$ (already shown in Eq. (\ref{q})):
\begin{eqnarray}
\frac{f/a}{m_P} &=& \frac{\sqrt{2}}{l} \,, \label{physpar1} \\
\frac{(f/a)'}{m_P} &=& \sqrt{2} \left( x - \frac{1}{l} \right) \,, \label{physpar2} \\
\frac{\phi}{m_P} &=& 2 \frac{w}{y l} \,, \label{physpar3} \\
\frac{\phi'}{m_P} &=& \sqrt{3} z \,. \label{physpar4}
\end{eqnarray}

\begin{itemize}
\item {\it First critical manifold}: \\
$w = 0, z = 0, v = 0, r = \sqrt{1 - x^2 - y^2}, l = 0.$ \\
For this manifold, $q = 1$, which means radiation domination.  This radiation can be pure or dark\footnote{By dark radiation we mean that associated to the kinetic term of the Yang-Mills fields.}, depending on the values of $x$ and $y$.  From the eigenvalues and eigenvectors, we can only conclude that the manifold is a saddle and that the Higgs field is evolving in a decelerated way since $z=0$ is a $z$-direction attractor.
\item {\it Second critical manifold}: \\
$y = 0, w = 0, z = 0, v = 0, r = \sqrt{1 - x^2}, l = 0.$ \\
This is a submanifold of the first one, for $y = 0$.  However, we can extract a bit more of information from it since $l = 0$ is, explicitly, a $l$-direction repeller.
For this manifold, $q = 1$, which means radiation domination.  This radiation can be pure or dark, depending on the value of $x$.  From the eigenvalues and eigenvectors, we can conclude that the manifold is a saddle, that the Higgs field is evolving in a decelerated way since $z=0$ is a $z$-direction attractor, and that the physical vector field starts with a large value but decreases in a decelerated way (in magnitude).
\item {\it Third critical manifold}: \\
$y = 0, w = 0, z = 0, v = 0, r = \sqrt{1 - x^2}, l = 1/x.$ \\
For this manifold, $q = 1$, which means radiation domination.  This radiation can be pure or dark, depending on the value of $x$.  From the eigenvalues and eigenvectors, we can conclude that the manifold is a saddle, that the Higgs field is evolving in a decelerated way since $z=0$ is a $z$-direction attractor, and that the physical vector field approaches a constant value, $\sqrt{2} x m_P$, because its speed approaches zero in view that $l = 1/x$ is a $l$-direction attractor.
\item {\it Fourth critical manifold}: \\
$y = \sqrt{1 - x^2}, w = 0, z = 0, v = 0, r = 0, l = 0.$ \\
For this manifold, $q = 1$, which means radiation domination.  This radiation cannot be pure on the manifold, but it can be pure or dark in its surroundings.  From the eigenvalues and eigenvectors, we can conclude that the manifold is a saddle and that the Higgs field is evolving in a decelerated way since $z=0$ is a $z$-direction attractor.
\item {\it First critical point}: \\
$x = 0, y = 0, w = 1, z = 0, v = 0, r = 0, l = 0.$ \\
For this point, $q = 0$, which means a transition from deceleration to acceleration.  From the eigenvalues and eigenvectors, we can conclude that the point is a saddle, that the Higgs field is evolving in a decelerated way since $z=0$ is a $z$-direction attractor, and that the radiation, pure and dark, is decreasing since $x = 0$, $y = 0$, and $r = 0$ are attractors in the $x$ direction, the $y$ direction, and the $r$ direction respectively.
\item {\it Second critical point}: \\
$x = 0, y = 0, w = 0, z = 1, v = 0, r = 0, l = 0.$ \\
For this point, $q = 2$, which means domination of the Higgs kinetic energy, what is called {\it ``kination''}.  From the eigenvalues and eigenvectors, we can conclude that the point is a repeller, that the Higgs field is growing in a decelerated way because, initially, $z=1$, it being a $z$-direction repeller, that the radiation, pure and dark, does not exist initially but it starts growing because, initially, $x = 0$, $y = 0$, and $r = 0$, they being repellers in the $x$ direction, the $y$ direction, and the $r$ direction respectively, and that the physical vector field starts with a large value because, initially, $l = 0$, it being a $l$-direction repeller, but decreases in a decelerated way (in magnitude) because, initially, $l = 0$.
\item {\it Third critical point}: \\
$x = 0, y = 0, w = 0, z = 0, v = 1, r = 0, l = 0.$ \\
For this point, $q = -1$, which means dark energy domination.  This, in turn, implies that the Hubble parameter is almost constant in the surroundings of the point.  Thus, from the eigenvalues and eigenvectors, we can conclude that the point is an attractor in the $\{ x, y, w, z, v, r \}$ space, which is what matters when calculating the deceleration parameter.  We can also conclude that the Higgs field approaches a constant value, because $v = 1$ is a $v$-direction attractor, but it does it in a decelerated way, because $z = 0$ is a $z$-direction attractor, that the radiation, pure and dark, approaches zero because $x = 0$, $y = 0$, and $r = 0$ are attractors in the $x$ direction, the $y$ direction, and the $r$ direction respectively, and that the physical vector field approaches zero, because $y = 0$ is a $y$-direction attractor, but it does it in a decelerated way (in magnitude) because $x = 0$ is a $x$-direction attractor while $l = 0$ is a $l$-direction repeller.
\item {\it Fourth critical point}: \\
$x = 1, y = 0, w = 0, z = 0, v = 0, r = 0, l = 0.$ \\
For this point, $q = 1$, which means radiation domination.  This radiation cannot be pure in the point but can be pure or dark in its surroundings.  From the eigenvalues and eigenvectors, we can conclude that the point is a saddle, that the Higgs field is evolving in a decelerated way since $z=0$ is a $z$-direction attractor, and that the physical vector field starts with a large value because, initially, $l = 0$, it being a $l$-direction repeller, but decreases in a decelerated way (in magnitude) because, initially, $l = 0$ and $x = 1$, the latter being a $x$-direction repeller.
\item {\it Fifth critical point}: \\
$x = 0, y = 0, w = 0, z = 0, v = 0, r = 0, l = 0.$ \\
For this point, $q = 1/2$, which means matter domination.  From the eigenvalues and eigenvectors, we can conclude that the point is a saddle, that the Higgs field is evolving in a decelerated way since $z=0$ is a $z$-direction attractor, that the radiation approaches zero because $x = 0$, $y = 0$, and $r = 0$ are attractors in the $x$ direction, the $y$ direction, and the $r$ direction respectively, and that the physical vector field starts with a large value because, initially, $l = 0$, it being a $l$-direction repeller, but decreases in a decelerated way (in magnitude) because, initially, $l = 0$.
\item {\it Sixth critical point}: \\
$x = -1, y = 0, w = 0, z = 0, v = 0, r = 0, l = -1.$ \\
For this point, $q = 1$, which means radiation domination.  From the eigenvalues and eigenvectors, we can conclude that the point is a saddle, that the Higgs field is evolving in a decelerated way since $z=0$ is a $z$-direction attractor, and that the physical vector field approaches $-\sqrt{2} m_P$ because $l = -1$, is a $l$-direction attractor.
\item {\it Seventh critical point}: \\
$x = 1, y = 0, w = 0, z = 0, v = 0, r = 0, l = 1.$ \\
For this point, $q = 1$, which means radiation domination.  From the eigenvalues and eigenvectors, we can conclude that the point is a saddle, that the Higgs field is evolving in a decelerated way since $z=0$ is a $z$-direction attractor, and that the physical vector field approaches $\sqrt{2} m_P$ because $l = 1$, is a $l$-direction attractor.
\end{itemize}

Notice that neither the critical points nor the stability criteria depend on the value of $\alpha$.  Moreover, it is clear from this analysis that, in the $\{ x, y, w, z, v, r \}$ space, the second critical point is the only repeller whereas the third critical point is the only attractor.

\section{Numerical solution} \label{numerical}

We will now proceed to present a numerical solution whose initial conditions\footnote{By initial conditions we mean the conditions today so the evolution during the past implies negative values for $N$.} are in agreement with the present observed values for the density parameters (see Refs. \cite{Aghanim:2018eyx,Lyth:2017hyk}).  We know that, today, the pure radiation density parameter defined as $\Omega_r \equiv \rho_r/3m_P^2 H^2$ has a value\footnote{The subindex 0 means that the corresponding quantity is evaluated today.} $\Omega_{r_0} \simeq 10^{-4}$ while the matter density parameter, defined as $\Omega_m \equiv \rho_m/3m_P^2 H^2$ has a value $\Omega_{m_0} \simeq 0.31$.  This implies the following initial values for the $r$ and $m$ dimensionless variables:
\begin{equation}
r_0 = 10^{-2} \,, \hspace{5mm} m_0 = 0.557 \,. \label{init1}
\end{equation}
We also know, from the previous section, that the final stage of the evolution of the Einstein Yang-Mills Higgs system is the dark energy dominated period (third critical point) where the energy budget is dominated by the Higgs potential energy, parameterized by the dimensionless variable $v$.  Thus, and having in mind the values and stability properties for all the variables in the third critical point and the constraint equation in Eq. (\ref{constraint}), we have chosen the following initial conditions for the other dimensionless variables that characterize the system under study:
\begin{eqnarray}
x_0 = 10^{-18} \,, && \hspace{5mm} y_0 = 10^{-18} \,, \nonumber \\
w_0 = 10^{-18} \,, && \hspace{5mm} z_0 = 10^{-18} \,, \nonumber \\
v_0 = 0.831 \,, &&  \hspace{5mm} l_0 = 10^2 \,. \label{init2}
\end{eqnarray}
Another motivation to choose these initial conditions has to do with the total length of the radiation dominated period\footnote{Assuming both, that the kination dominated period may be removed and that the reheating period is intantaneous.}, from the end of the kination dominated period to the time of matter-radiation equality:
\begin{equation}
\Delta N_{\rm rad} \approx  - \frac{1}{2} \ln \left( \frac{\sqrt{\Omega_{m_0}} H_0 (1 + z_{r_{\rm eq}})^{3/2}}{H_{\rm end}} \right) \,, \label{lengthrad}
\end{equation}  
where $z_{r_{\rm eq}}$ is the redshift at matter-radiation equality and $H_{\rm end}$ is the Hubble parameter at the end of the kination dominated period.  Since nucleosynthesis has to occur during the radiation dominated period, $H_{\rm end} \gsim 10^{-45} m_P$.  Additionally, the constraint on the tensor to scalar ratio \cite{Akrami:2018odb} gives the upper bound $H_{\rm end} \lsim 10^{-2} m_P$ when the reheating is instantaneous and the length of the kination dominated period is considered to vanish.  These two bounds, together with Eq. (\ref{lengthrad}), lead to $13 \lsim \Delta N_{\rm rad} \lsim 62$.  As we will see, the above initial conditions lead to a cosmological dynamics consistent with the latter constraint.  The way Eq. (\ref{lengthrad}) is obtained is presented in Appendix \ref{applengthrad}.

\begin{figure}
\includegraphics[height=5cm,width=0.8\linewidth]{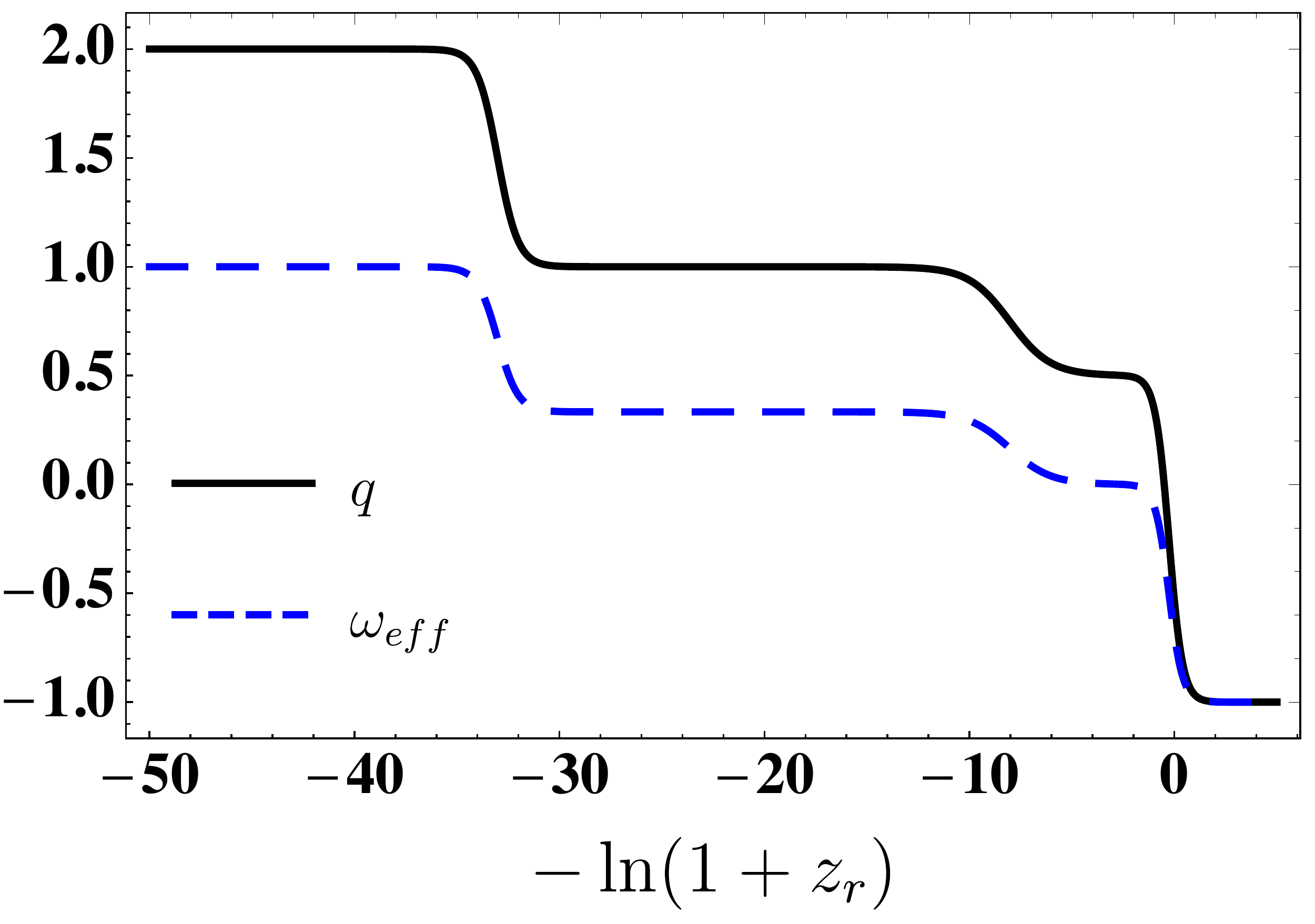}%
\caption{Evolution of the deceleration parameter $q$ and the effective equation of state parameter $\omega_{\rm eff}$.  The time variable is given by the number of e-folds of expansion $N$ which is equal to $-\ln (1 + z_r)$, with $z_r$ being the redshift.  It is easy to appreciate four stages of evolution:  the kination dominated period from early times to $N \simeq -32$, the radiation dominated period from $N \simeq -32$ to $N \simeq -8$, the matter dominated period from $N \simeq -8$ to $N \simeq -0.3$, and the dark energy dominated period from $N \simeq -0.3$ onwards.  With the realistic initial conditions given by Eqs. (\ref{init1})-(\ref{init2}), the system does not approach enough to the first critical point (transition from deceleration to acceleration).}
\label{qandomega}
\end{figure}

\begin{figure}
\includegraphics[height=5cm,width=0.8\linewidth]{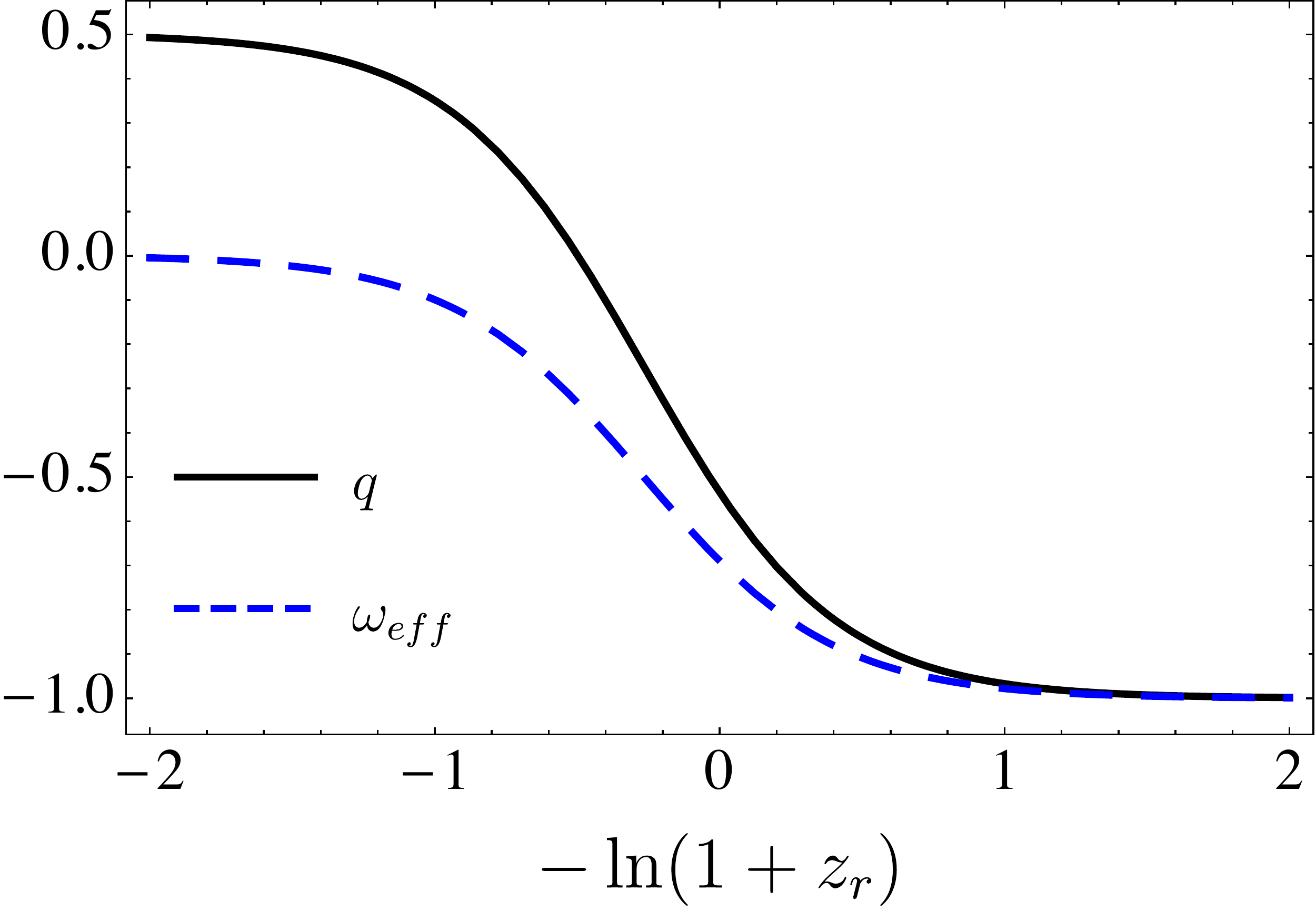}%
\caption{Zoom of Figure \ref{qandomega} around the present time $N=0$.  $\omega_{\rm eff} \neq -1$ for $N=0$.}
\label{qandomegazoom}
\end{figure}

\begin{figure}
\includegraphics[height=5cm,width=0.8\linewidth]{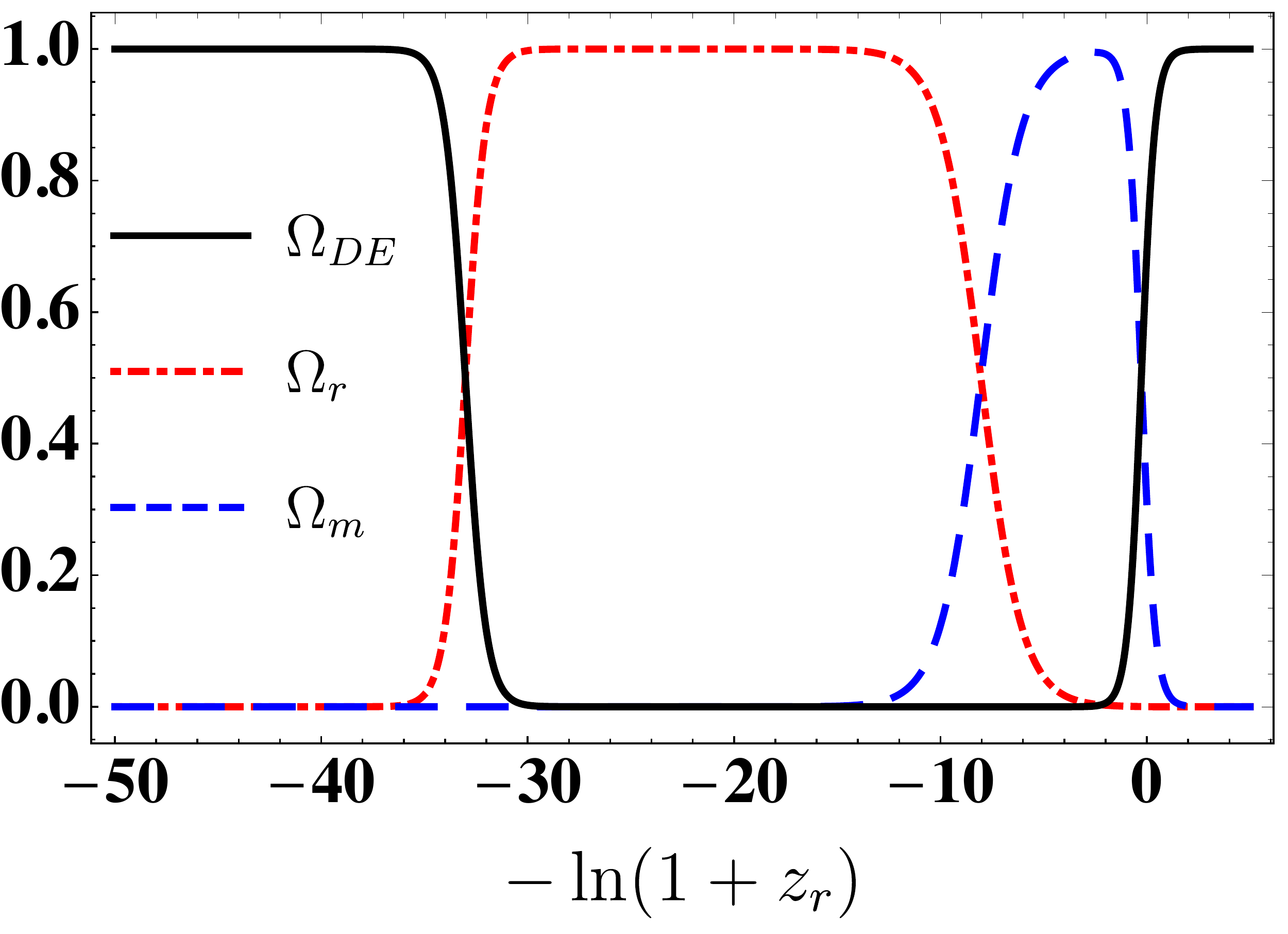}%
\caption{Evolution of the density parameters for radiation, $\Omega_r$, matter, $\Omega_m$, and dark energy $\Omega_{\rm DE}$.  The dark components (Yang-Mills + Higgs) support the kination dominated period in early times as well as the dark energy dominated period in the present and future.}
\label{densities}
\end{figure}

We have numerically solved the system of equations given by Eqs. (\ref{xeq})-(\ref{leq}) assuming a realistic value for the $\alpha$ parameter, $\alpha = 1$ \footnote{This is the order of magnitude obtained when calculating $\alpha$ in the SM.}. 
This allows us to plot the evolution of the deceleration parameter $q$ and the effective equation of state parameter $\omega_{\rm eff}$ given by $\omega_{\rm eff} = (2q -1)/3$;  such plots are shown in Figs. \ref{qandomega} and \ref{qandomegazoom}.  As observed, the system exhibits four stages of evolution:  the kination dominated period, the radiation dominated period, the matter dominated period, and, finally, the dark energy dominated period.  Except for the first one, the other periods and their relative order are in agreement with the standard cosmology (see Ref. \cite{Lyth:2017hyk});  nevertheless, a kination dominated period previous to the radiation dominated one is absolutely beneficial for a successful reheating mechanism if the inflaton is only coupled to matter gravitationally \cite{Opferkuch:2019zbd}.  Regarding the qualitative behaviour discussed in the previous section, we can conclude that, with the realistic initial conditions given by Eqs. (\ref{init1})-(\ref{init2}), the system does not approach enough to the first critical point (transition from deceleration to acceleration).  We can also confirm that the kination dominated period is an attractor towards the past (a repeller or source), the radiation and matter dominated periods are metastable (saddle points), and the dark energy dominated period is an attractor (a sink).  The evolution of the density parameters for radiation, $\Omega_r$, for matter, $\Omega_m$, and for dark energy, $\Omega_{\rm DE}$, this latter defined as $\Omega_{\rm DE} \equiv 1 - \Omega_r - \Omega_m$, is presented in Fig. \ref{densities}. As we can see, the dark components (Yang-Mills + Higgs) support the kination dominated period in early times as well as the dark energy dominated period in the present and future, similarly to what happens with most of the quintessence models \cite{Amendola:2015ksp}.  

We have checked that the changes in Figs. (\ref{qandomega}) and (\ref{densities}) are negligible when increasing or decreasing $x_0$ and $y_0$.  However, an increment in the value of $w_0$ does modify the figures, shortening the radiation dominated period while lengthening the kination dominated one without modifying the length of the matter dominated period; a similar behaviour has been found for changes in $z_0$, $l_0$, or $\alpha$.  It is remarkable that the behaviour in these plots is consistent with the redshift at the matter-radiation equality, $z_{r_{\rm eq}} = 3390$ \cite{Amendola:2015ksp,Aghanim:2018eyx,Lyth:2017hyk};  it is also consistent with the constraint on the length of the radiation dominated period given below Eq. (\ref{lengthrad}).  Let's analyze now each one of the relevant periods discussed above.

\subsection{The kination dominated period}

\begin{figure}
\subfloat[\label{higgskination}]{%
  \includegraphics[height=3.5cm,width=.48\linewidth]{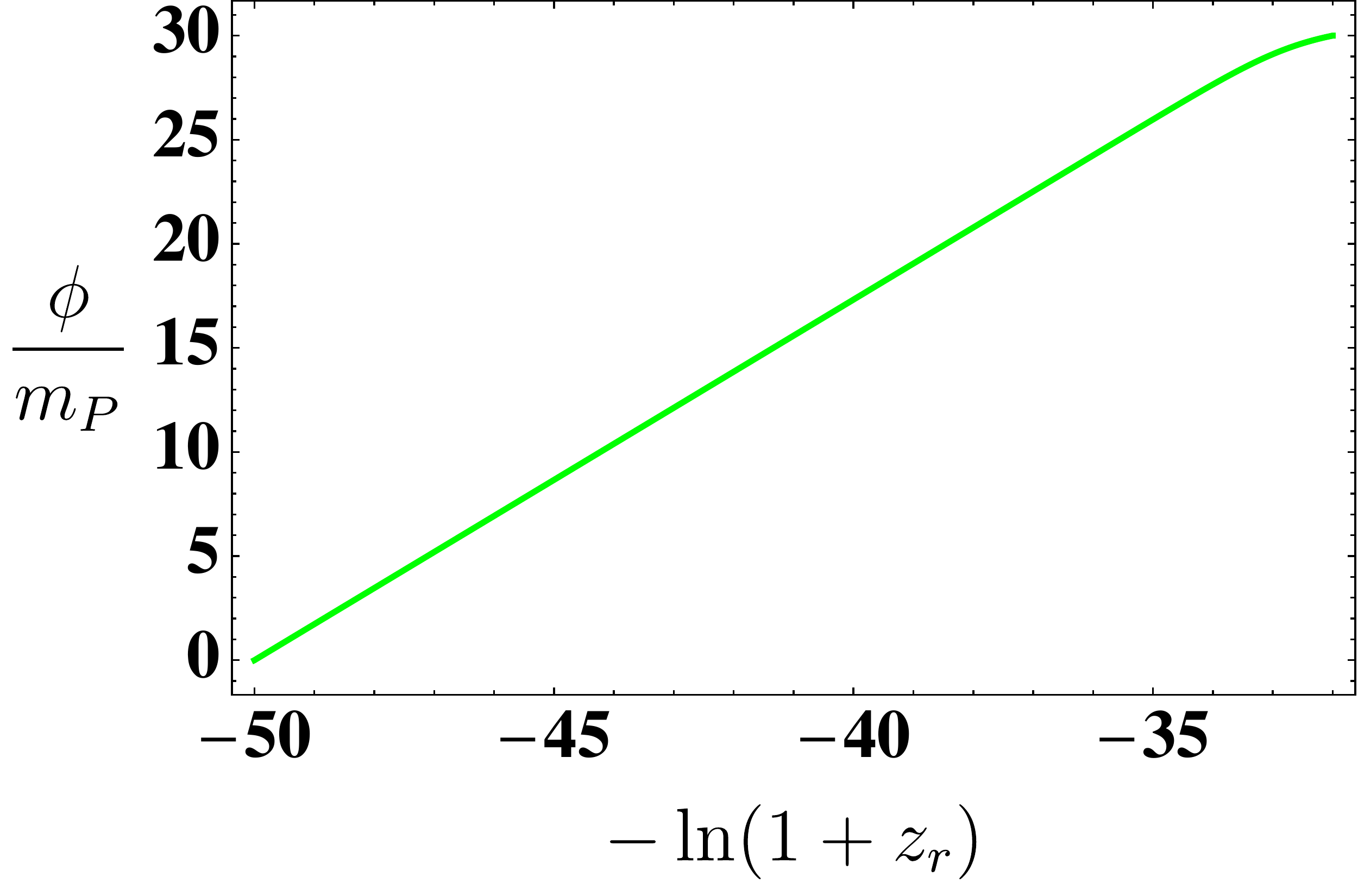}%
}\hfill
\subfloat[\label{higgsspeedkination}]{%
  \includegraphics[height=3.5cm,width=.48\linewidth]{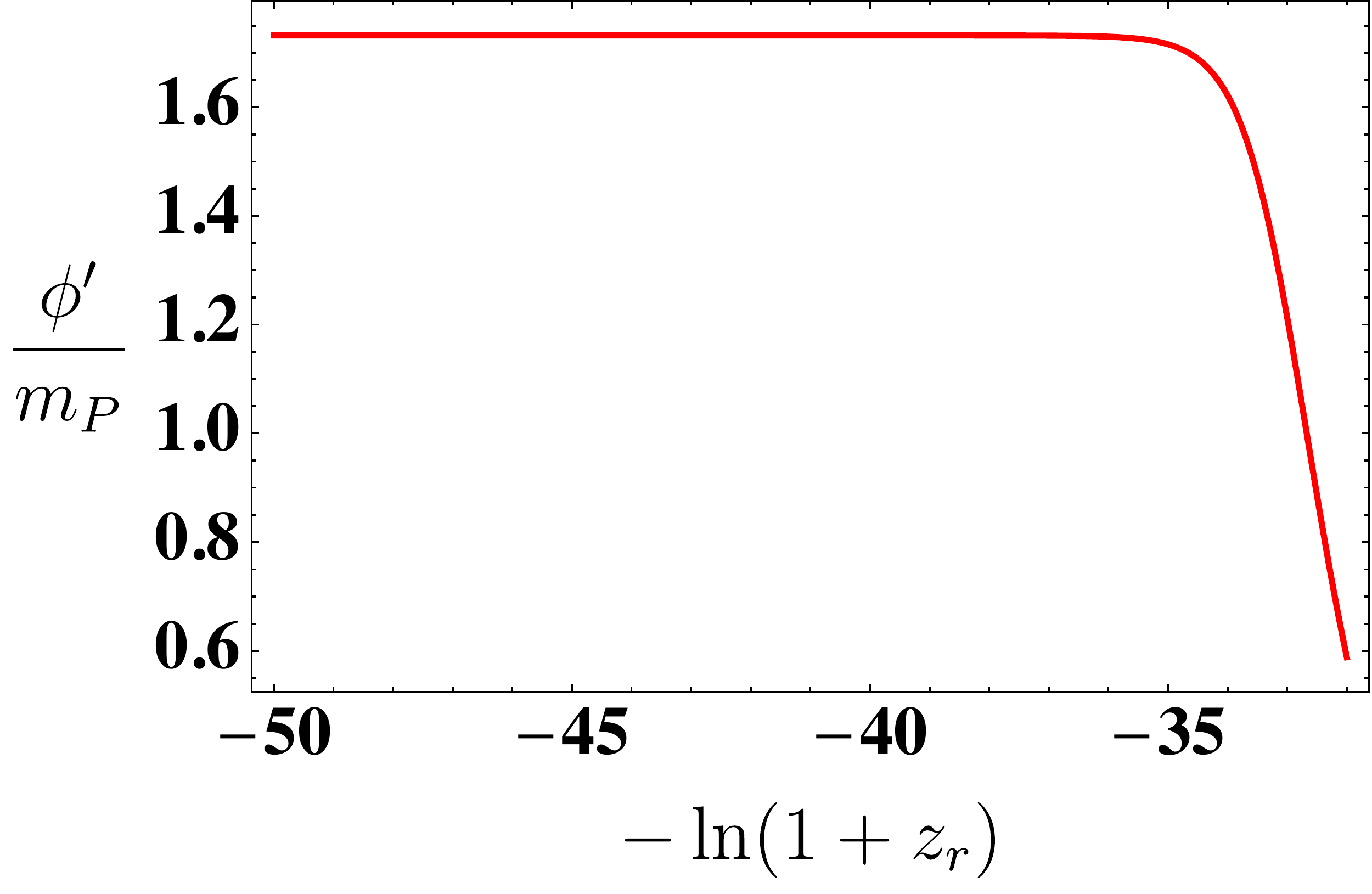}%
}\\
\subfloat[\label{vectorkination}]{%
  \includegraphics[height=3.5cm,width=.48\linewidth]{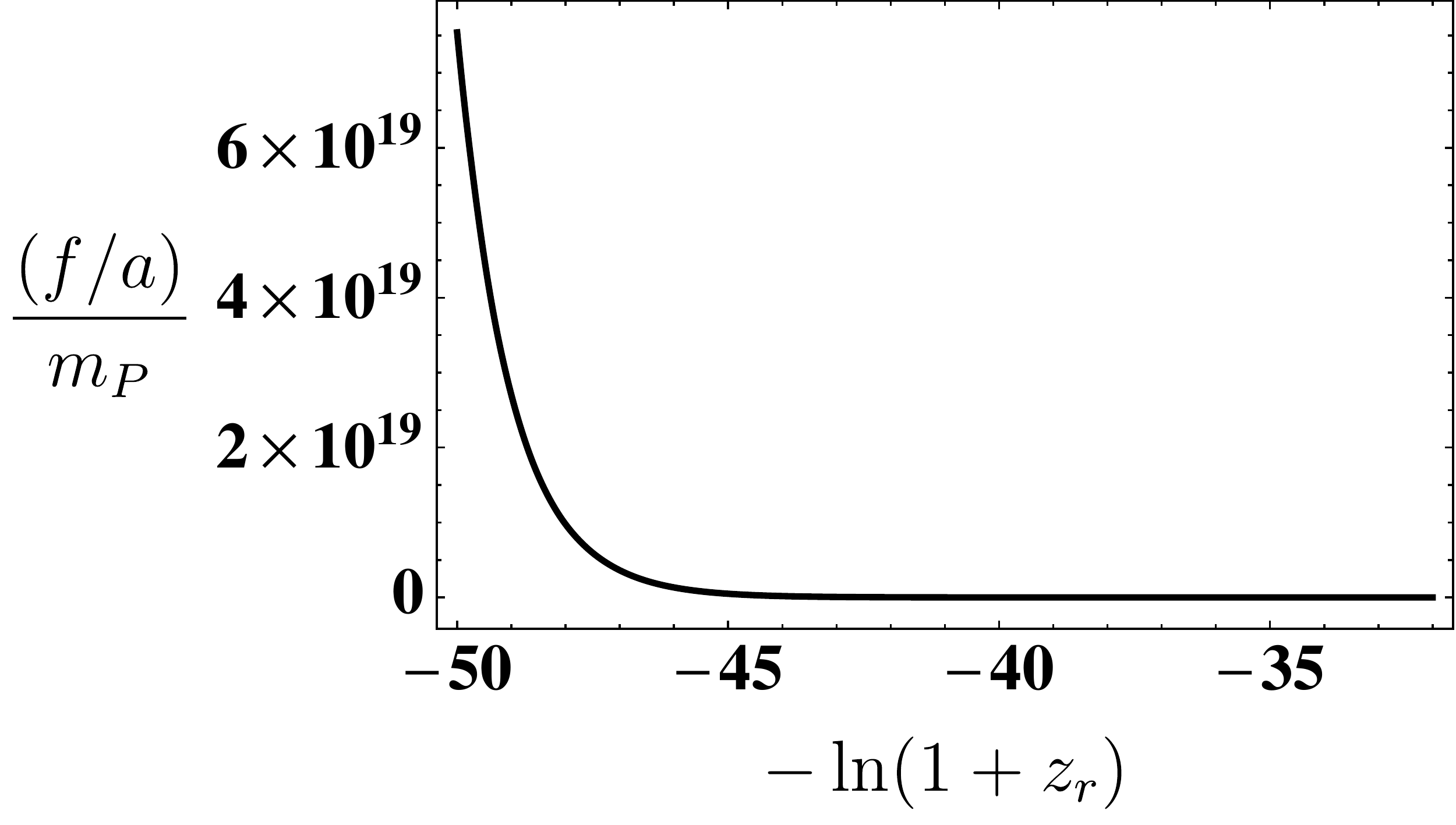}%
}\hfill  
\subfloat[\label{vectorspeedkination}]{%
  \includegraphics[height=3.5cm,width=.48\linewidth]{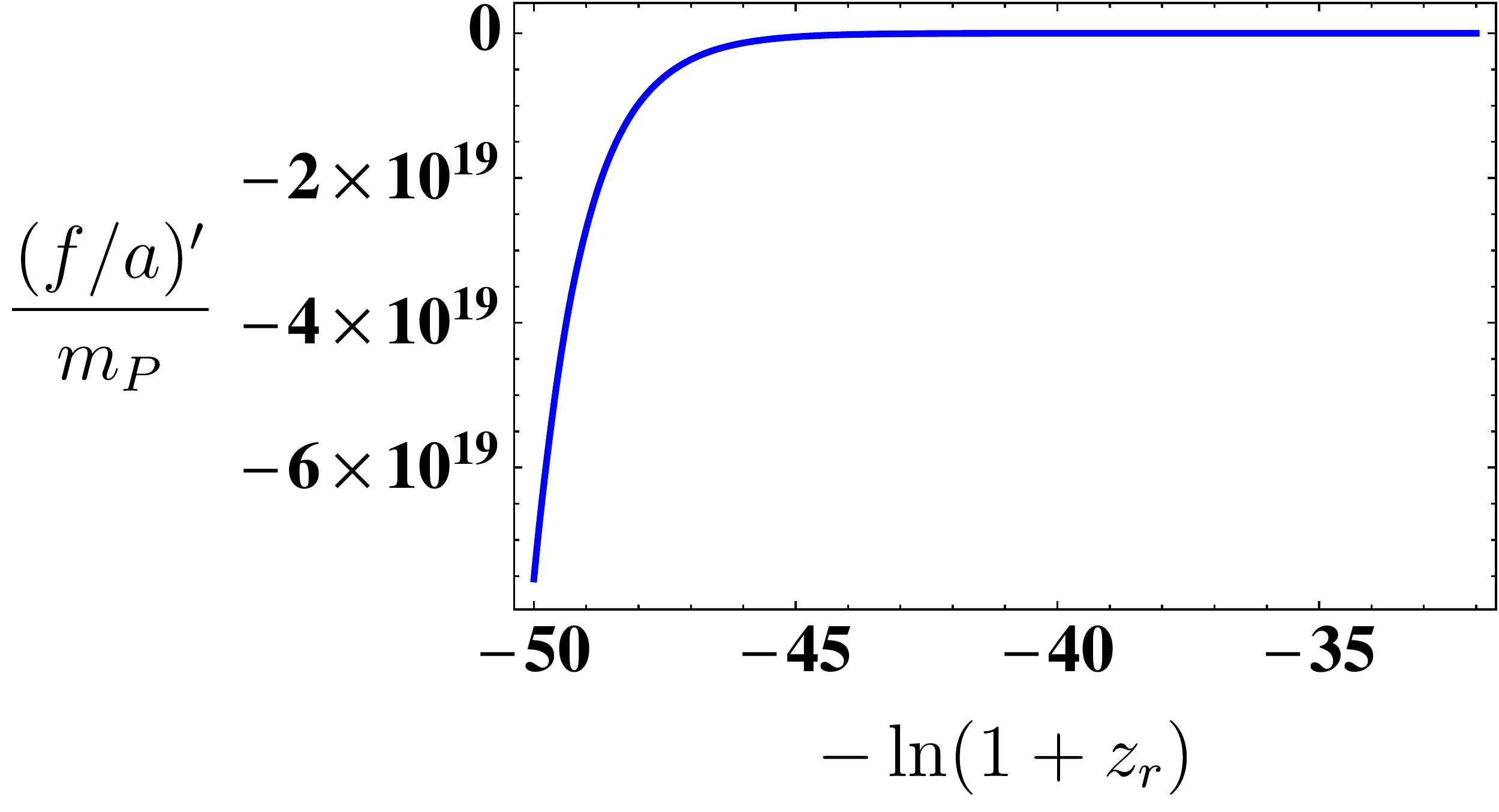}%
}   
\caption{Evolution during the kination dominated period (second critical point).  Plots a), b), c), and d) show the Higgs field, its speed, the physical vector field, and its speed respectively.  The behaviour is in agreement with that obtained and described in Section \ref{dynsys} via dynamical systems.}
\label{physvarkin}
\end{figure}

This period runs from early times to $N \simeq -32$ and corresponds to the second critical point.  The evolution of the physical variables presented in Eqs. (\ref{physpar1}) - (\ref{physpar4}) is shown in Fig. \ref{physvarkin}.  The Higgs field grows in a decelerated way while the vector field decreases from a huge value in a decelereated way (in magnitude).  Such behaviour is in agreement with that obtained and described in the previous section.

\subsection{The radiation dominated period}

\begin{figure}
\subfloat[\label{higgsradiation}]{%
  \includegraphics[height=3.5cm,width=.48\linewidth]{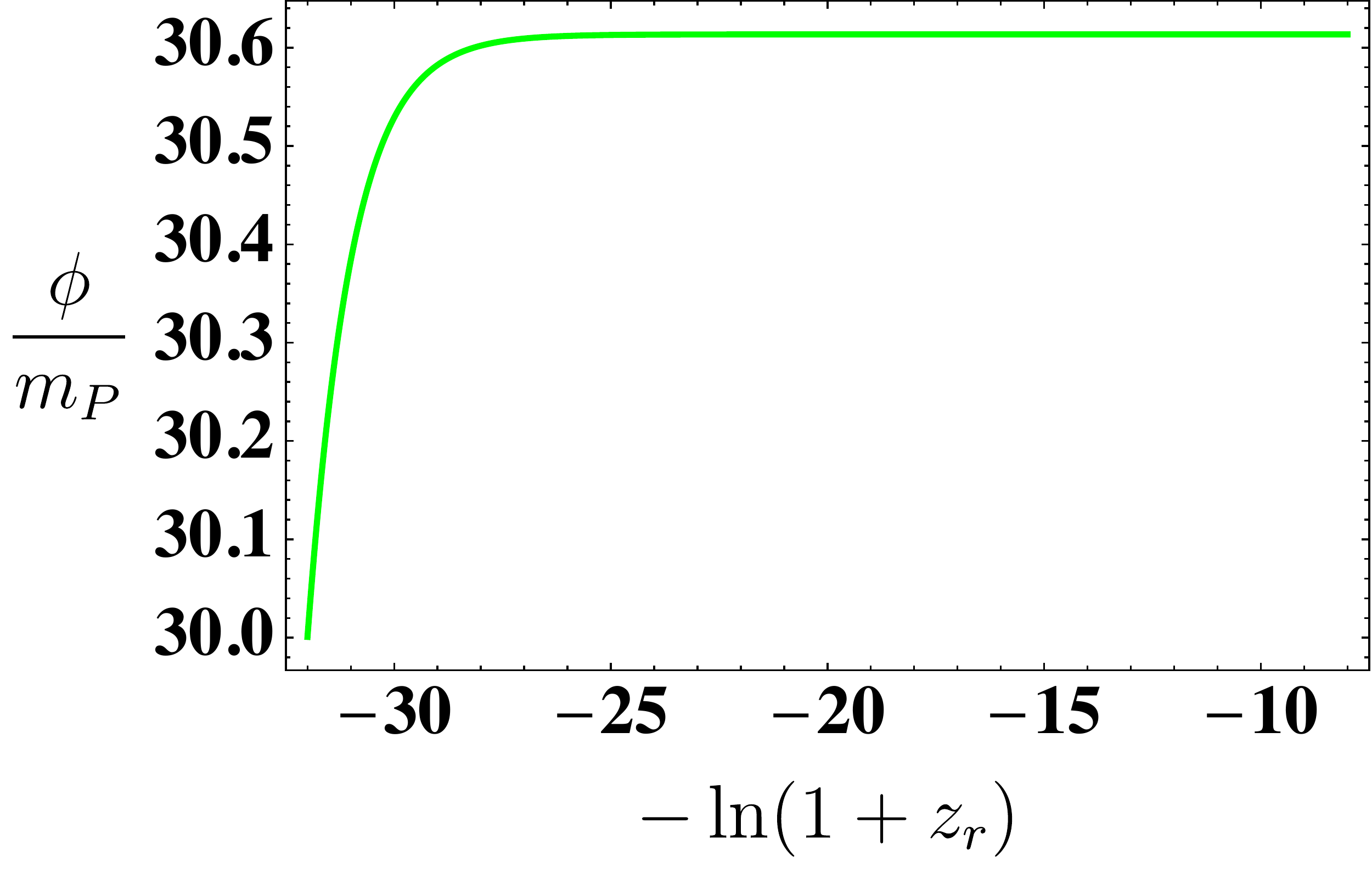}%
}\hfill
\subfloat[\label{higgsspeedradiation}]{%
  \includegraphics[height=3.5cm,width=.48\linewidth]{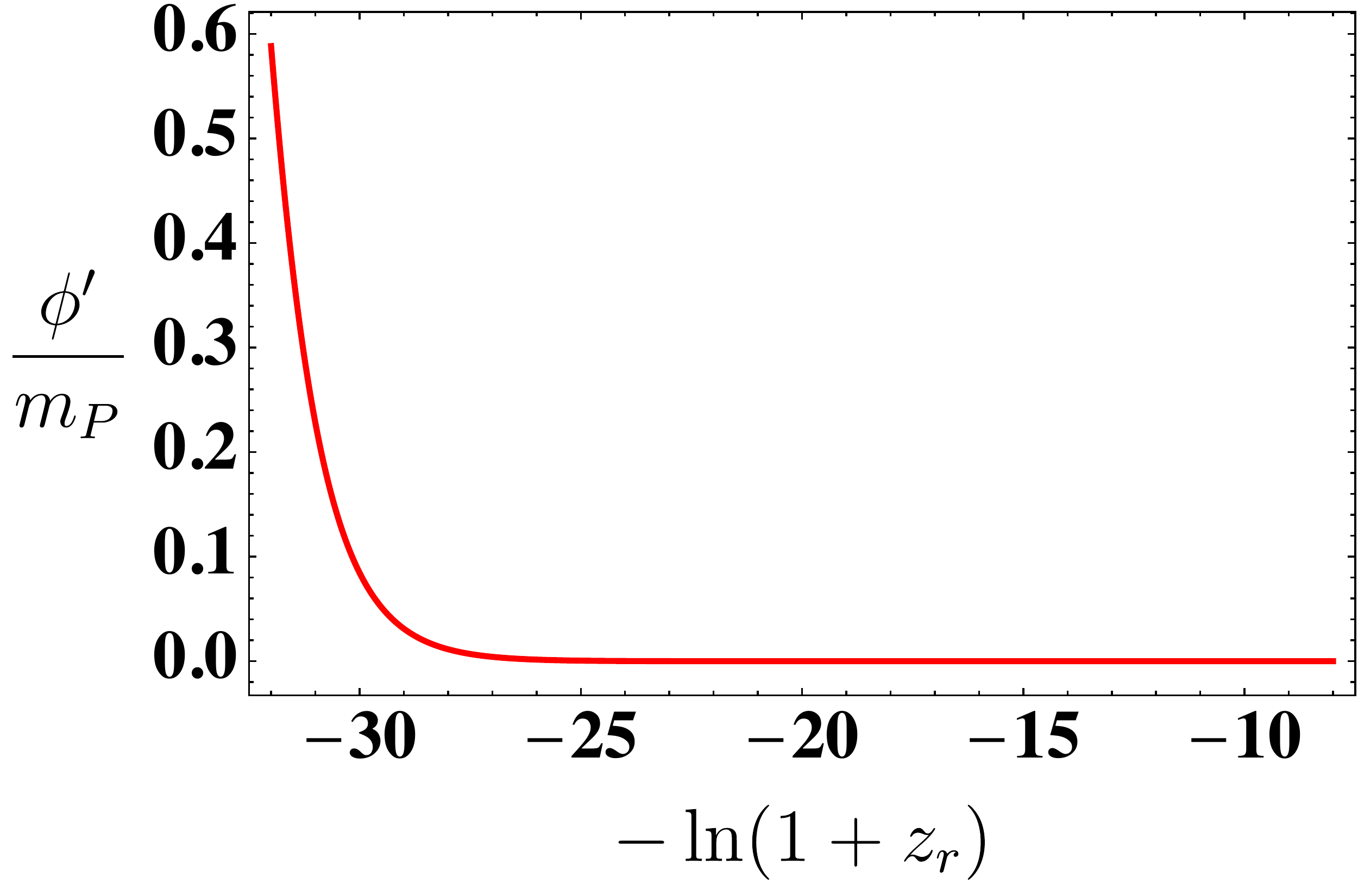}%
}\\
\subfloat[\label{vectorradiation}]{%
  \includegraphics[height=3.5cm,width=.48\linewidth]{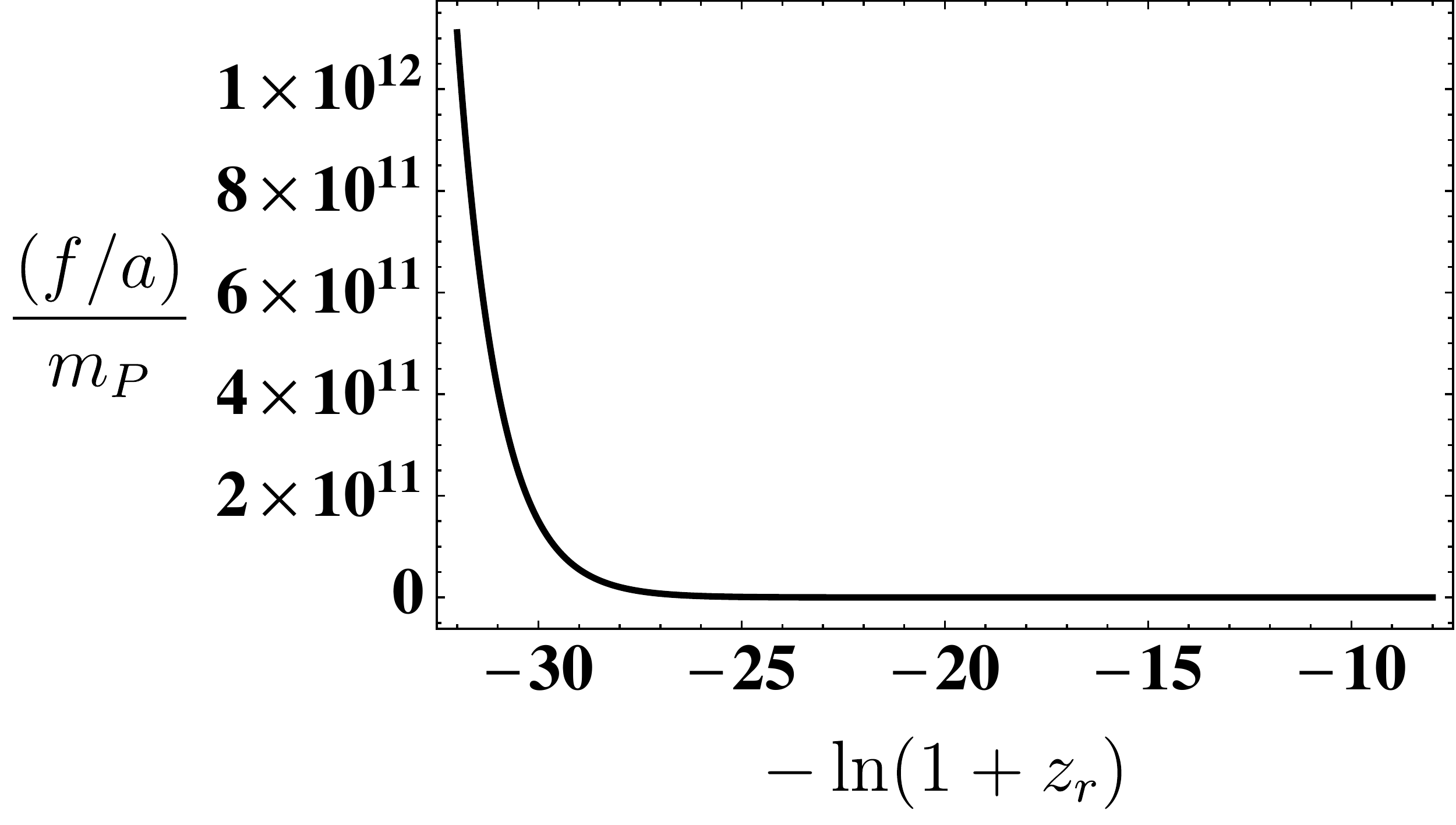}%
}\hfill  
\subfloat[\label{vectorspeedradiation}]{%
  \includegraphics[height=3.5cm,width=.48\linewidth]{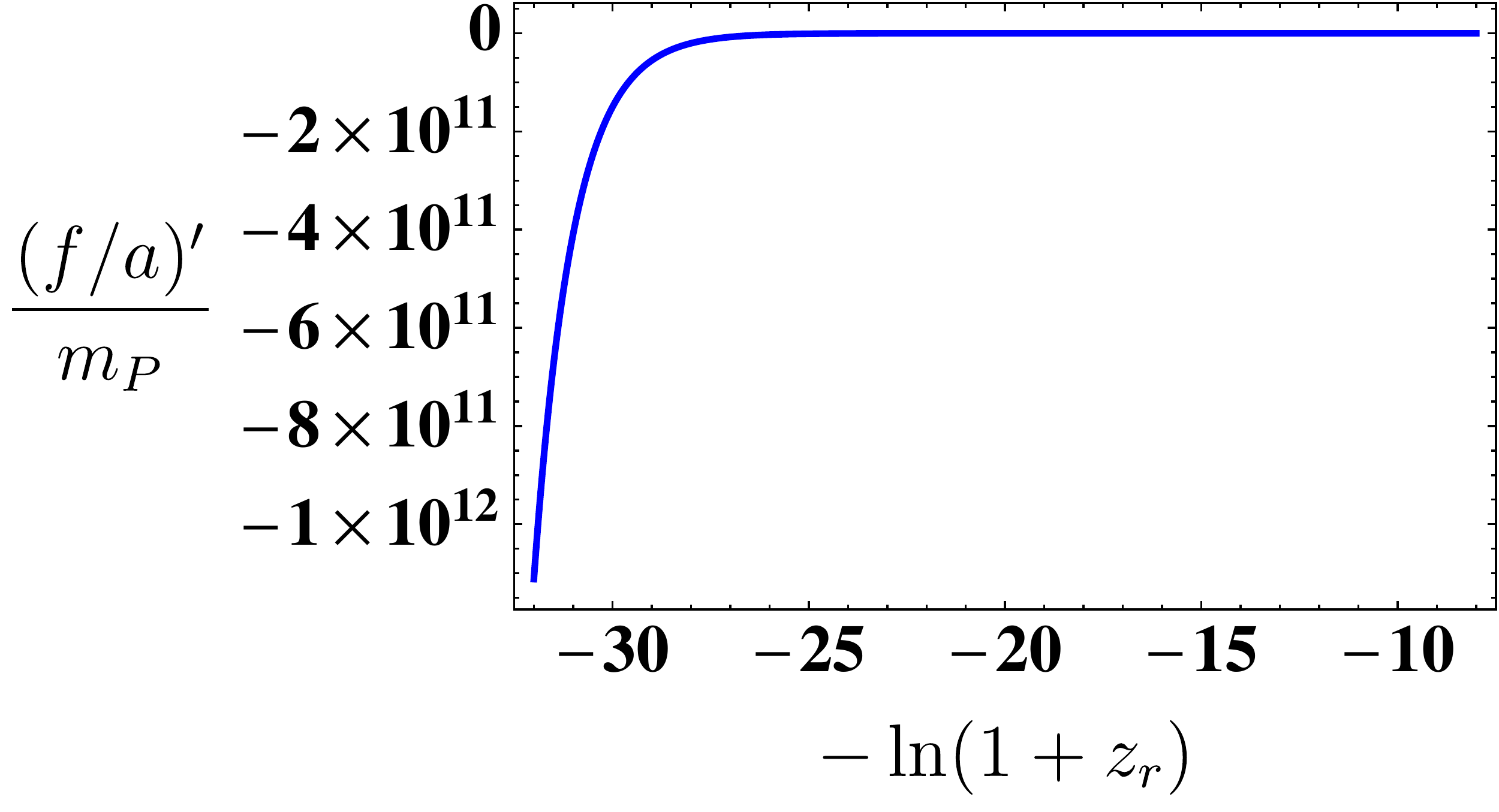}%
}   
\caption{Evolution during the radiation dominated period (first or second critical manifold).  Plots  a), b), c), and d) show the Higgs field, its speed, the physical vector field, and its speed respectively.  The behaviour is in agreement with that obtained and described in Section \ref{dynsys} via dynamical systems.}
\label{physvarrad}
\end{figure}

This period runs from $N \simeq -32$ to $N \simeq -8$, i.e. $\Delta N_{\rm rad} \simeq 24$, and corresponds to either the first or second critical manifold.  The evolution of the physical variables presented in Eqs. (\ref{physpar1}) - (\ref{physpar4}) is shown in Fig. \ref{physvarrad}.  As in the kination dominated period, the Higgs field grows in a decelerated way;  regarding the physical vector field, it continues decreasing in a decelereated way (in magnitude).  Such behaviour is in agreement with that obtained and described in the previous section.

\subsection{The matter dominated period}

\begin{figure}
\subfloat[\label{higgsspeedmatter}]{%
  \includegraphics[height=3.5cm,width=.48\linewidth]{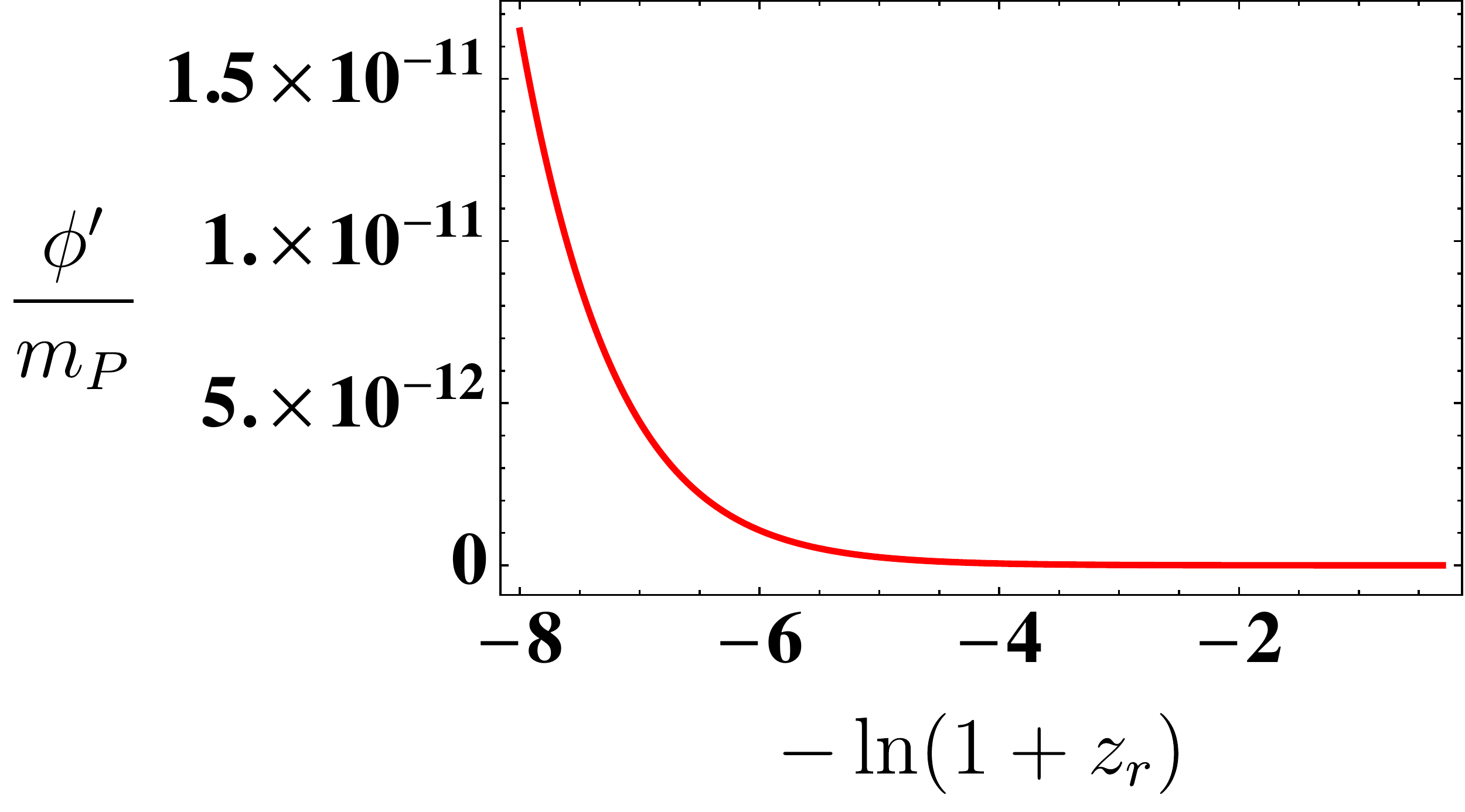}%
}\\
\subfloat[\label{vectormatter}]{%
  \includegraphics[height=3.5cm,width=.48\linewidth]{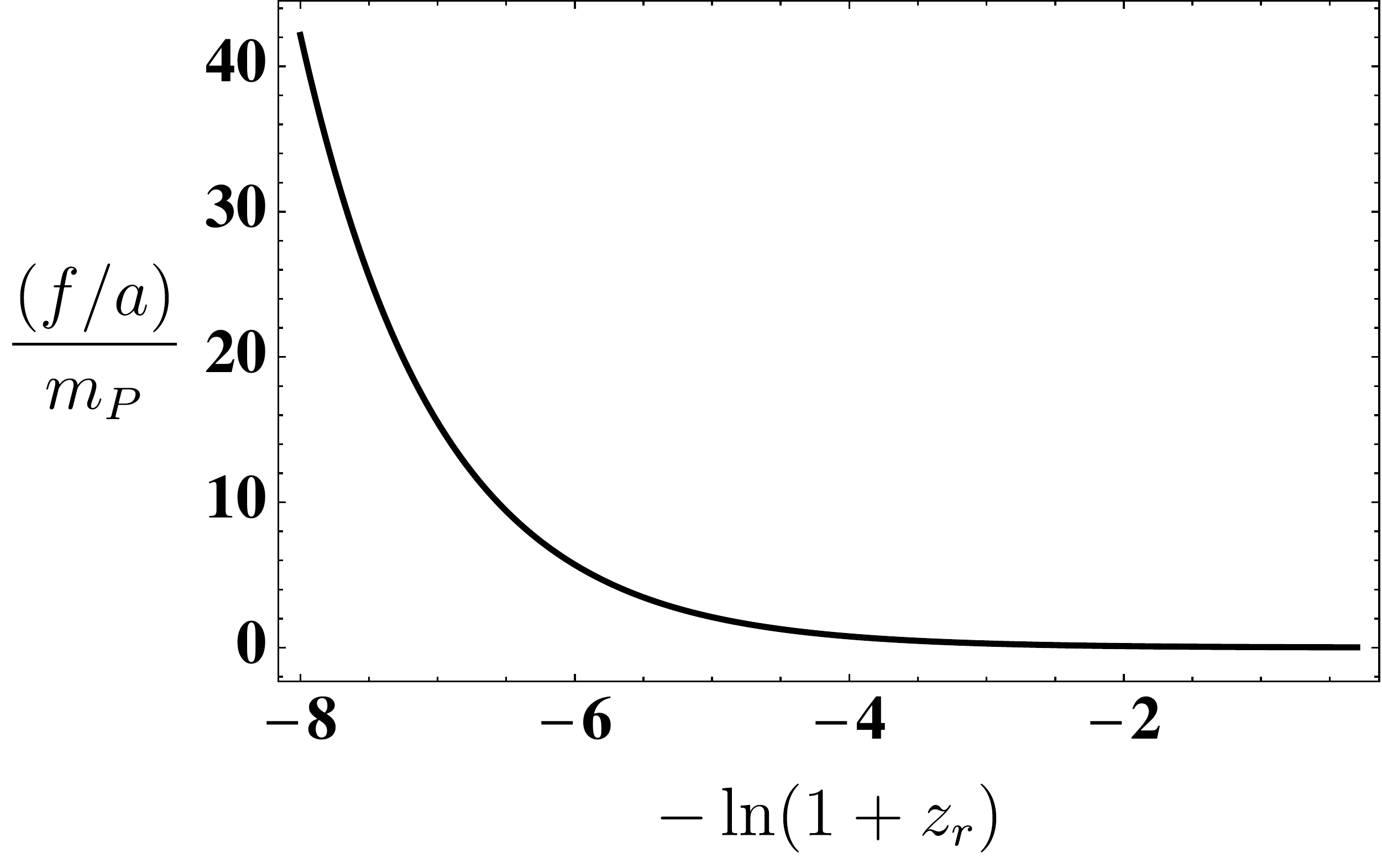}%
}\hfill  
\subfloat[\label{vectorspeedmatter}]{%
  \includegraphics[height=3.5cm,width=.48\linewidth]{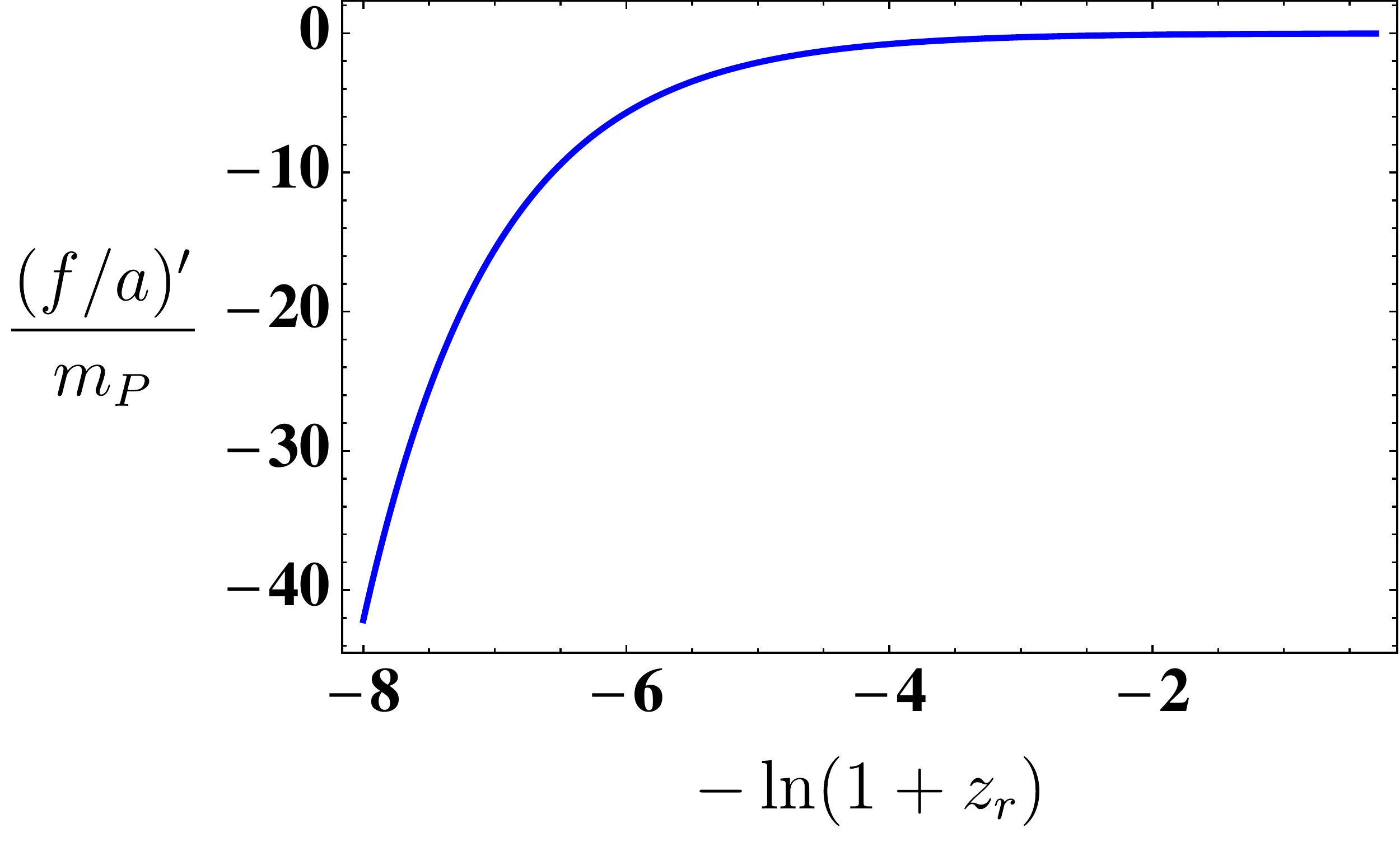}%
}   
\caption{Evolution during the matter dominated period (fifth critical point).  Plots a), b), and c), show the Higgs speed, the physical vector field, and its speed respectively. The evolution of the Higgs field is not presented since, by this stage, it has almost reached its asymptotic value. The behaviour is in agreement with that obtained and described in Section \ref{dynsys} via dynamical systems.}
\label{physvarmat}
\end{figure}

This period runs from $N \simeq -8$ to $N \simeq -0.3$ and corresponds to the fifth critical point.  The evolution of the physical variables presented in Eq. (\ref{physpar1}) and Eqs. (\ref{physpar3}) - (\ref{physpar4}) is shown in Fig. \ref{physvarmat}.  The Higgs field still grows in a decelerated way but very slowly because it has almost reached its asymptotic value (that is the reason why the Higgs field plot is not shown in Fig. \ref{physvarmat}).  In the meantime, the vector field continues decreasing in a decelereated way (in magnitude).  Such behaviour is in agreement with that obtained and described in the previous section.

\subsection{The dark energy dominated period}

\begin{figure}
\subfloat[\label{higgsspeeddark}]{%
  \includegraphics[height=3.5cm,width=.48\linewidth]{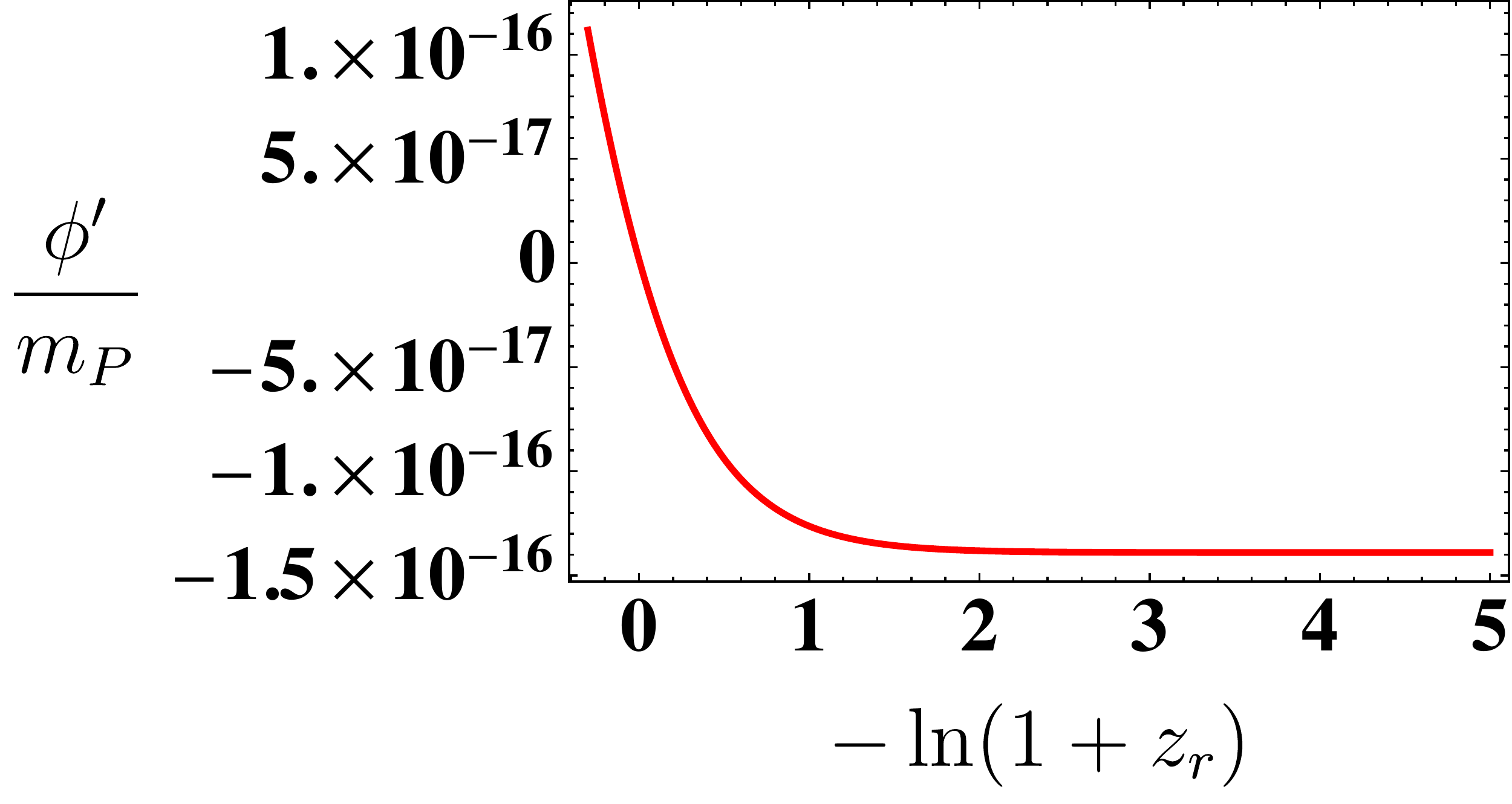}%
}\\
\subfloat[\label{vectordark}]{%
  \includegraphics[height=3.5cm,width=.48\linewidth]{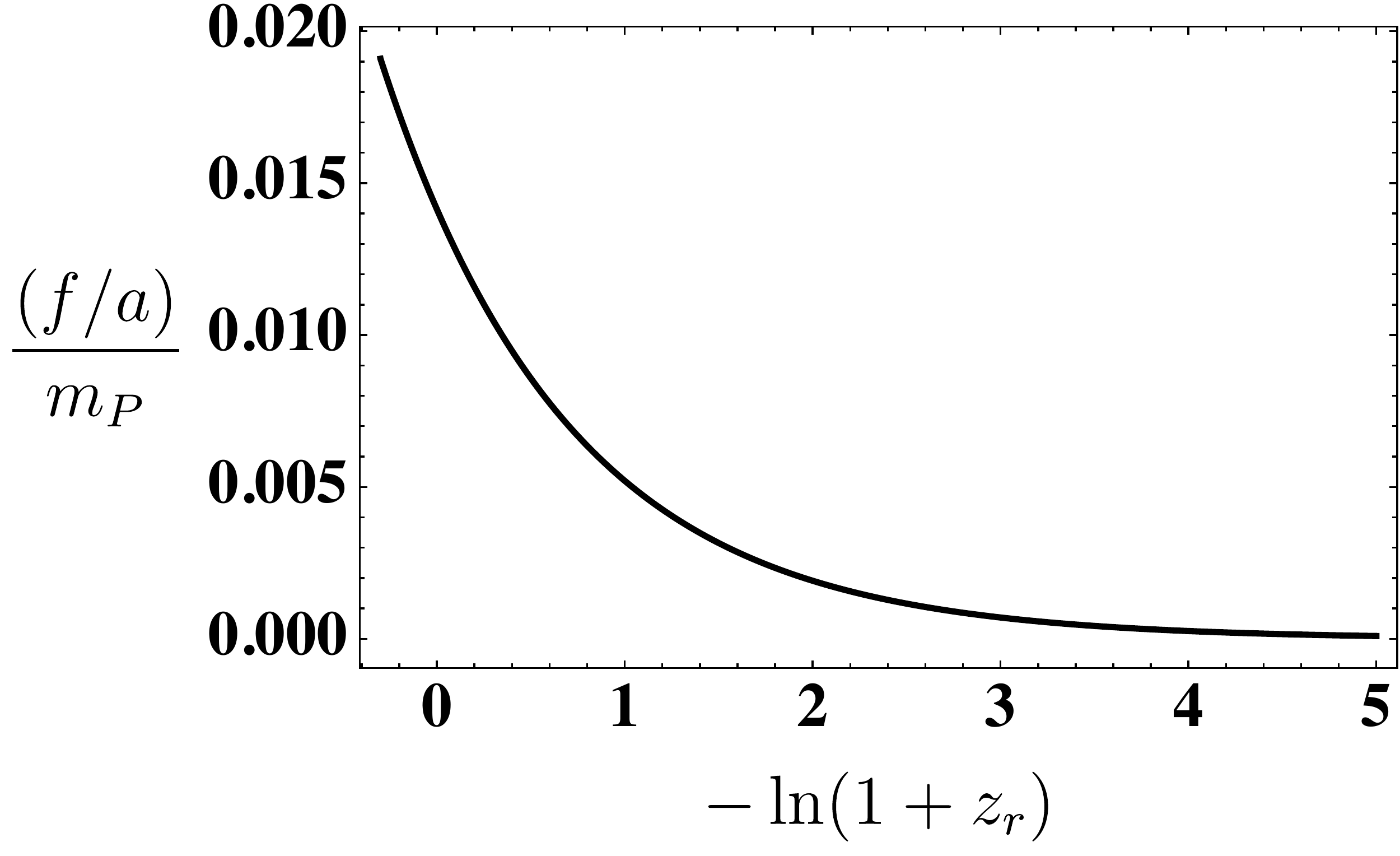}%
}\hfill  
\subfloat[\label{vectorspeeddark}]{%
  \includegraphics[height=3.5cm,width=.48\linewidth]{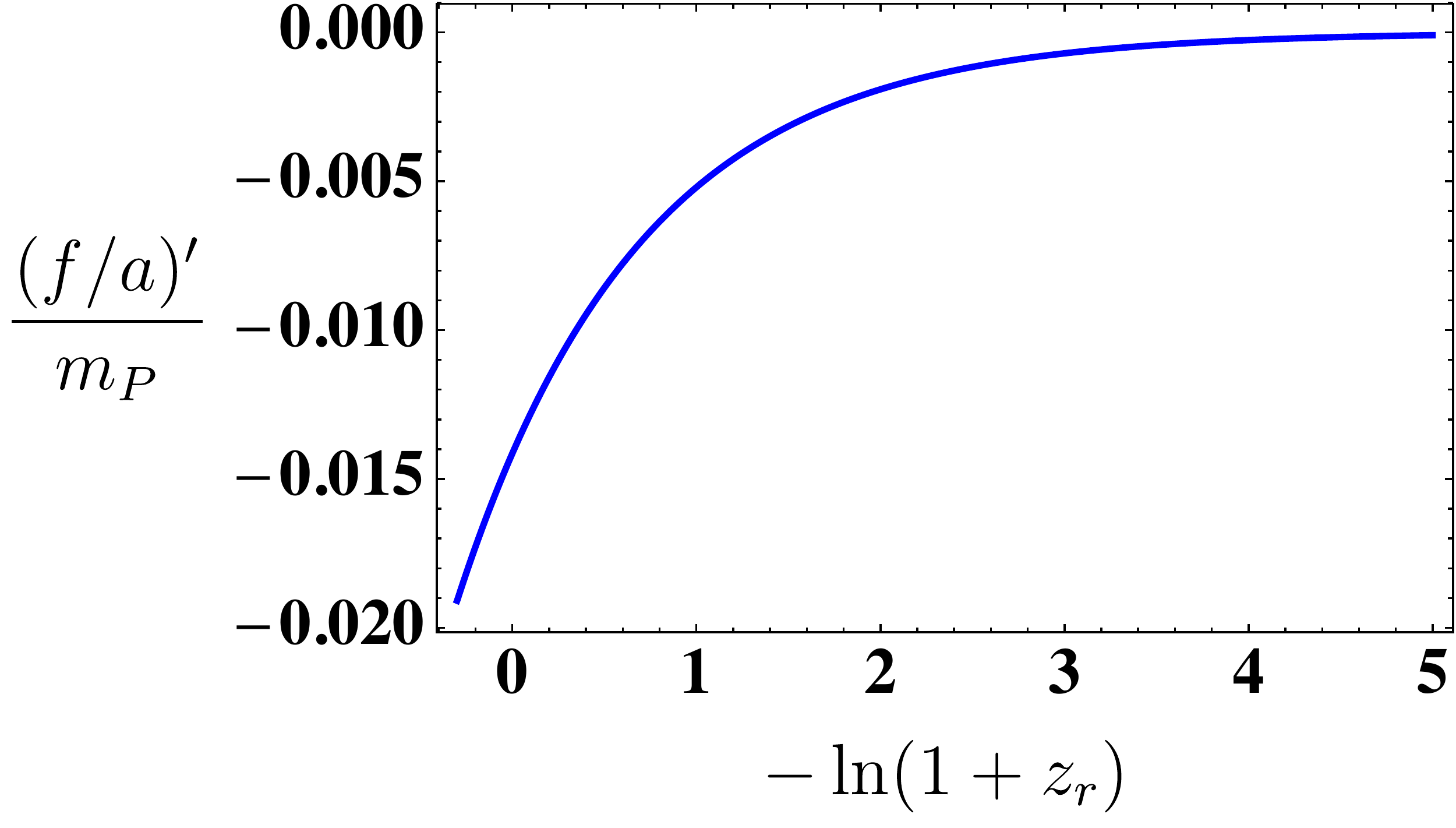}%
}   
\caption{Evolution during the dark energy dominated period (third critical point).  Plots a), b), and c), show the Higgs speed, the physical vector field, and its speed respectively.  The behaviour is in agreement with that obtained and described in Section \ref{dynsys} via dynamical systems.}
\label{physvardark}
\end{figure}

This period runs from $N \simeq -0.3$ onwards and corresponds to the third critical point.  The evolution of the physical variables presented in Eq. (\ref{physpar1}) and Eqs. (\ref{physpar3}) - (\ref{physpar4}) is shown in Fig. \ref{physvardark}.  The Higgs field has almost reached its asymptotic value, still growing in a decelerated way while the vector field continues decreasing in a decelereated way (in magnitude) towards its asymptotic value (zero).  Such behaviour is in agreement with that obtained and described in the previous section.

\subsection{Global evolution}

\begin{figure}
\subfloat[\label{higgsglobal}]{%
  \includegraphics[height=3.5cm,width=.48\linewidth]{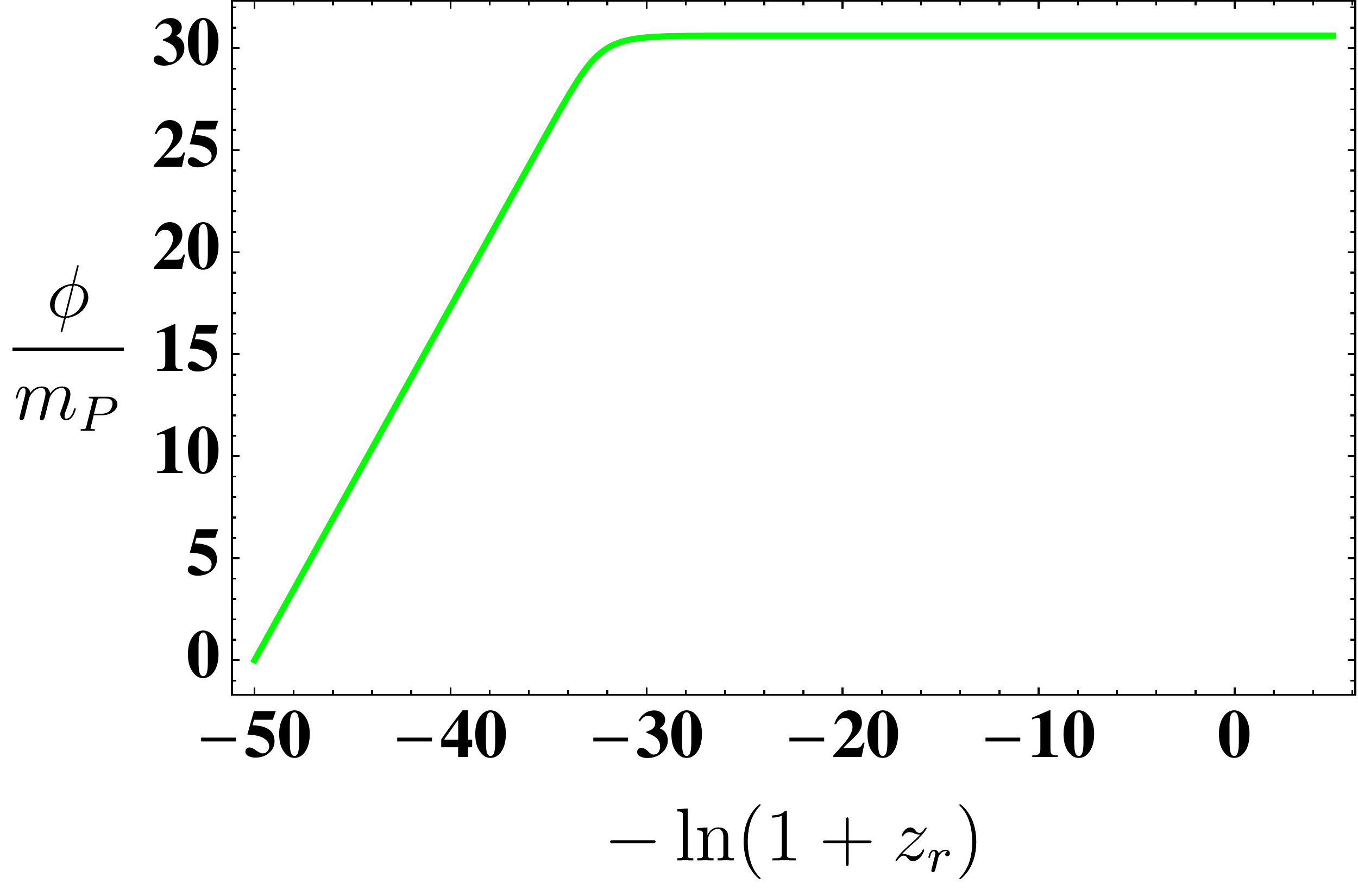}%
}\hfill
\subfloat[\label{higgsspeedglobal}]{%
  \includegraphics[height=3.5cm,width=.48\linewidth]{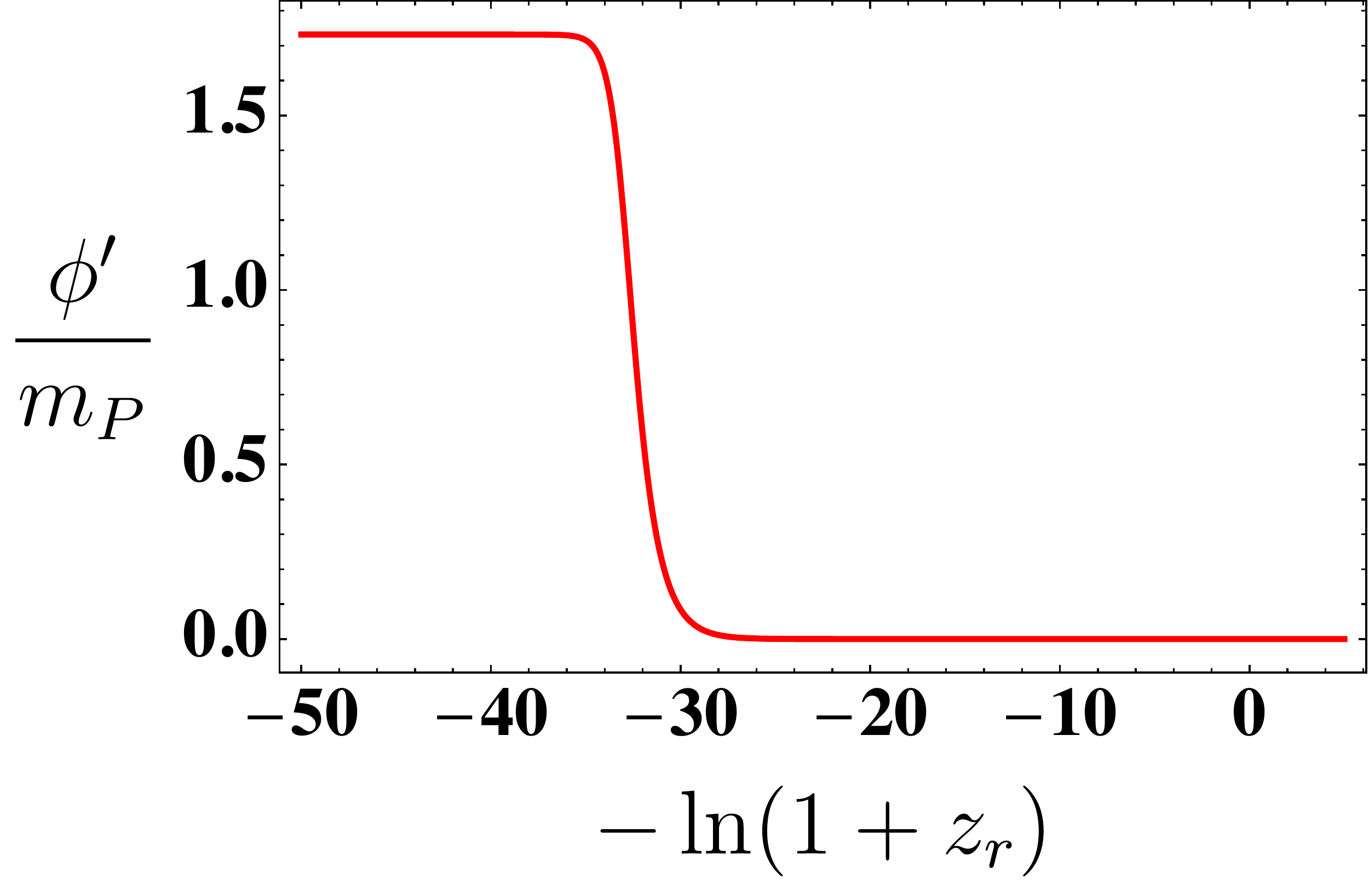}%
}\\
\subfloat[\label{vectorglobal}]{%
  \includegraphics[height=3.5cm,width=.48\linewidth]{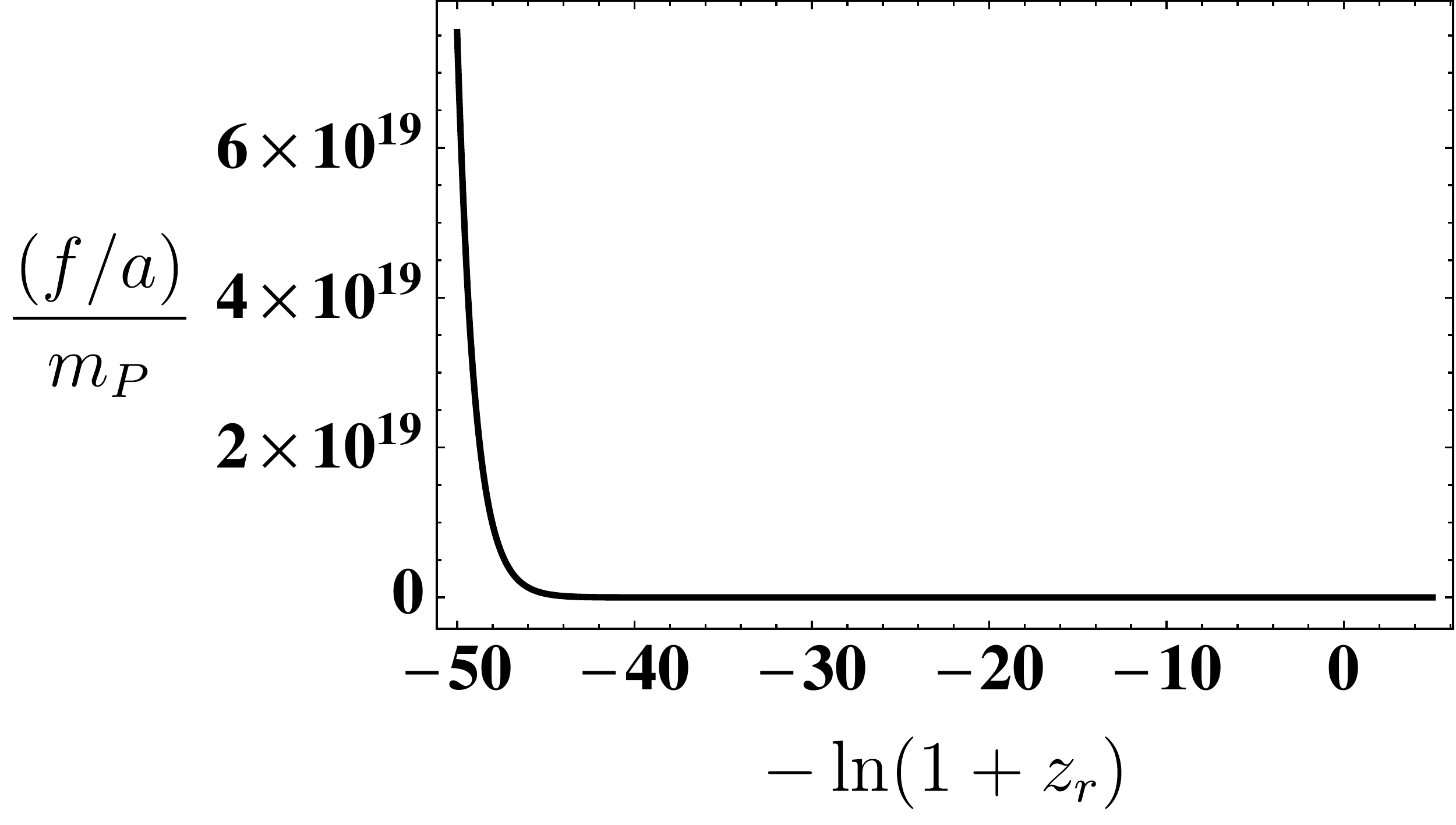}%
}\hfill  
\subfloat[\label{vectorspeedglobal}]{%
  \includegraphics[height=3.5cm,width=.48\linewidth]{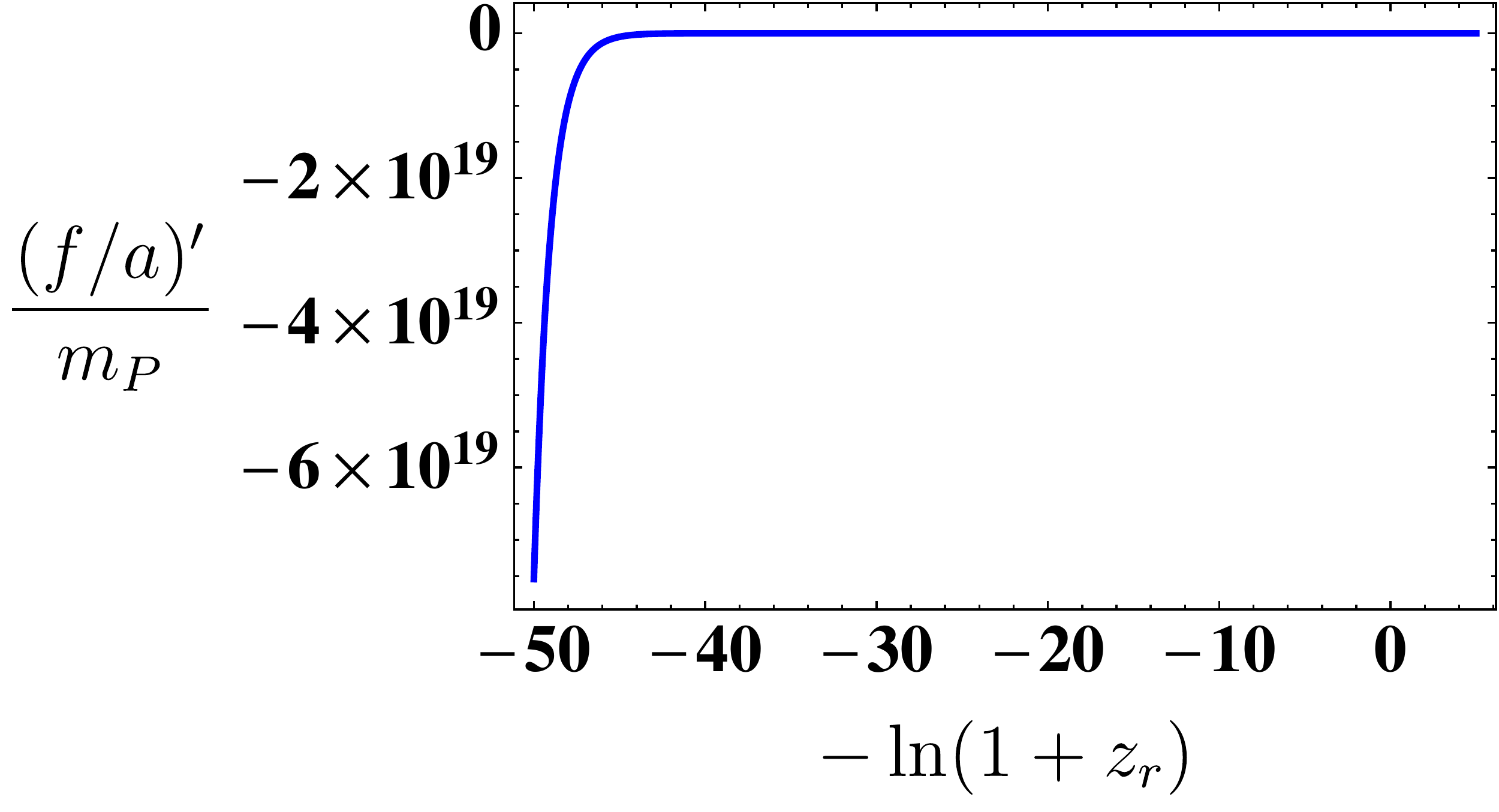}%
}   
\caption{Global evolution of the Einstein Yang-Mills Higgs system.  Plots a), b), c), and d) show the Higgs field, its speed, the physical vector field, and its speed respectively.  The Higgs field moves away from the minimum of its potential, being hold towards an asymptotic finite value by its interaction with the Yang-Mills fields.  The physical vector field, which represents the Yang-Mills fields, starts from a huge value and decreases towards zero, it being its asymptotic value.  It is the interaction between these two types of fields that lifts the Higgs field and keeps it away from its minimum.  This, in turn, produces the dark energy dominated period.}
\label{global}
\end{figure}

We can take a global look at the evolution of the physical quantities described in Eqs. (\ref{physpar1}) - (\ref{physpar4}) (see Fig. \ref{global}). We can conclude that {\em the Higgs field moves away from the minimum of its potential towards an asymptotic finite value, being hold by its interaction with the Yang-Mills fields which do not allow it to roll down the Mexican hat potential.} (see Fig. \ref{higgsglobal});  of course, since it is approaching an asymptotic value, its movement is decelerated (see Fig. \ref{higgsspeedglobal}).  In contrast, {\em the physical vector field, which represents the Yang-Mills fields, starts from a huge value and decreases towards zero, it being its asymptotic value} (see Fig. \ref{vectorglobal});  its movement, of course, is decelerated (in magnitude) (see Fig. \ref{vectorspeedglobal}).

It is worth mentioning that if the cosmic triad were actually zero, the interaction term in Eq. (\ref{phieq}) would disappear and the Higgs field would go to the minimum of its potential, which is indeed the behaviour, for instance, in the SM;  no late accelerated expansion period would be generated in this case.  However, when recognizing that the gauge field does have a potential, and that it rolls down towards the minimum of this potential, the interaction between the gauge field and the Higgs field always survives because the evolution for $f/a$ becomes asymptotic banning $f/a$ to reach zero (see Fig. \ref{vectorglobal}) and approaching $\phi$ to a non-vanishing asymptotic value (see Fig. \ref{higgsglobal}).  This asymptotic behaviour reveals that the accelerated expansion period is eternal which is welcome, although not necessary, in a dark energy scenario.

\section{Observational signatures} \label{signatures}

\begin{figure}
\includegraphics[height=5cm,width=0.8\linewidth]{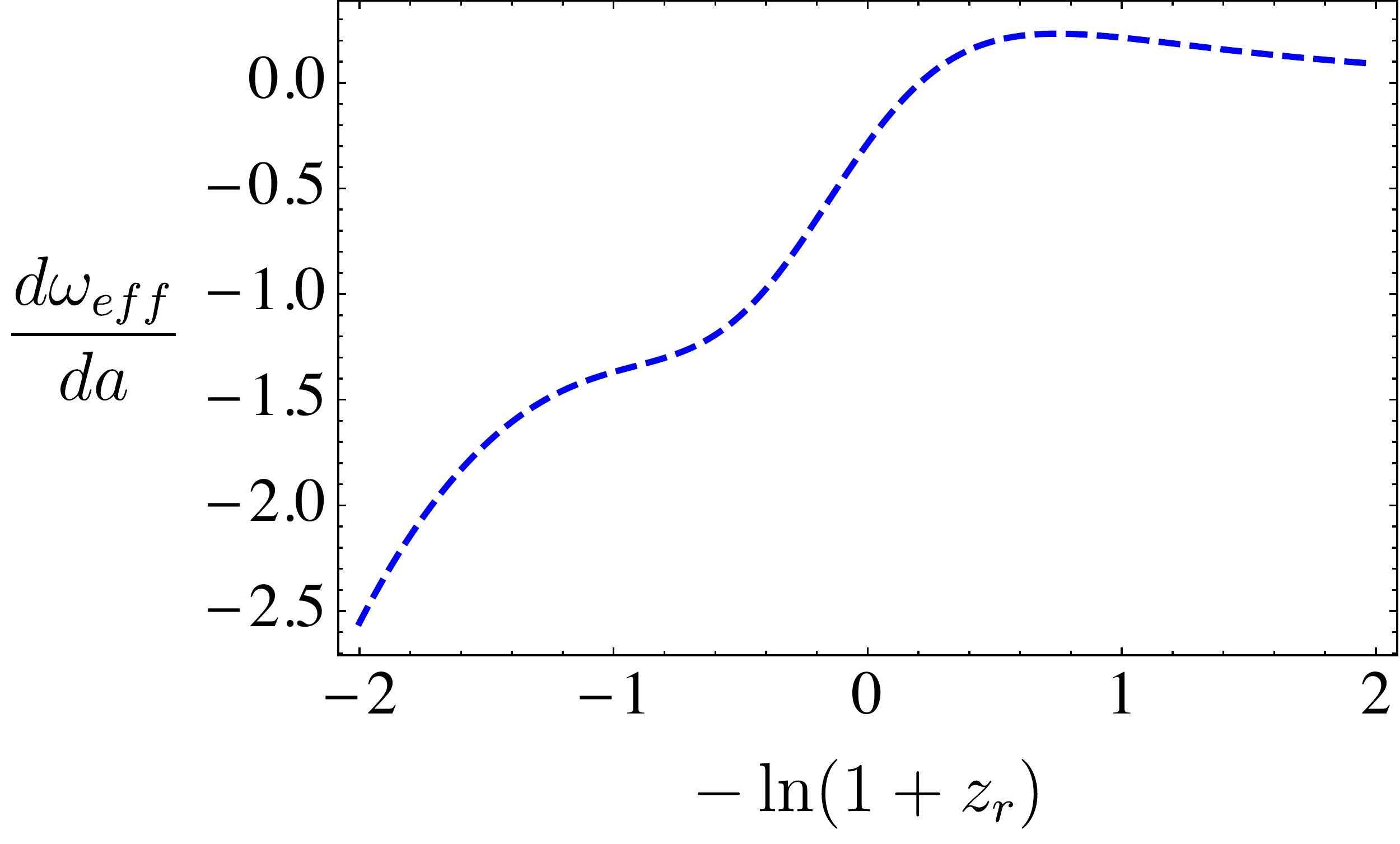}%
\caption{$d \omega_{\rm eff}/ d a$ as a function of $-\ln(1+z_r)$.  This plot reveals that $\omega_{\rm eff}$ does not change linearly with the expansion parameter $a$, so it is not possible to employ the CPL parametrization \cite{Chevallier:2000qy,Linder:2002et}.}
\label{omegaevolution}
\end{figure}

Dark energy models can be distinguished through both its background evolution and its effects on the growth of perturbations in matter.  While the former is tested by distance measurements, the latter is tested by cosmic microwave background anisotropies, weak lensing, and the distribution of large-scale structures \cite{Amendola:2015ksp,Rajvanshi:2019wmw}.  As seen in Figs. (\ref{qandomega}) and (\ref{qandomegazoom}), the time evolution of $\omega_{\rm eff}$ for the Einstein Yang-Mills Higgs model clearly differs from that of a cosmological constant.  The characteristics of such time evolution are better appreciated in Fig. (\ref{omegaevolution}) which reveals that $\omega_{\rm eff}$ does not evolve linearly with the expansion parameter $a$, so it is not possible to employ the Chevallier-Polarski-Linder (CPL) parametrization $\omega_{\rm eff} = p + q(a - a_i) $ \cite{Chevallier:2000qy,Linder:2002et} where $p$ and $q$ are constants and $a_i$ is the expansion parameter when $\omega_{\rm eff} = p$.  Comparison with distance measurements requires certain parametrization for the time evolution of $\omega_{\rm eff}$, the cosmological constant and the CPL parametrization being the most employed;  however, these do not work for the model studied and, therefore, a suitable measurement analysis for this model, which goes beyond the scope of this paper, must be carried out.  Now, regarding the effects on the growth of perturbations in matter, a complete study of the first-order cosmological perturbations is required in order to see the possible deviations in the effective gravitational coupling and the sound speed for each mode (see e.g. Ref. \cite{Kase:2018aps}).  On the other hand, we know that vector fields are source of perturbations that introduce violations of the isotropy, which directly affect the growth of matter perturbations and, therefore, become relevant to the matter or galaxy spectrum or bispectum \cite{Bartolo:2017sbu,Bonvin:2017req,MoradinezhadDizgah:2018pfo,Tansella:2018hdm}. New observables are, therefore, introduced and become a window to test the Einstein Yang-Mills Higss model. For example, it is possible to  identify two explicit contributions from vector fields to the two-point correlation function of the galaxy distribution \cite{Bonvin:2017req}. The first one is related to the two-point correlation function of vector fields perturbations $\langle \delta A_i\ \delta A_j \rangle$, which gives us information about the amplitude of fields and  statistical anisotropy patterns \cite{Bonvin:2017req,Tansella:2018hdm}. The other one is the cross-correlation with the matter component $\langle\delta A_i \ \delta M\rangle$ which is a signal of  statistical anisotropy.  Both contributions are directly related to the parameters of the present model and could be measured by forthcoming experiments designed to map the large-scale structure \cite{Bartolo:2017sbu,Bonvin:2017req,Tansella:2018hdm}. To uncover the specific relation between the matter correlation functions and the specific parameters in the Einstein Yang-Mills Higgs scenario, we need to go beyond the simple background analysis and perform a deep one in the frame of cosmological perturbation theory. Such study also goes beyond the scope of this paper.

\section{Some implications for particle physics} \label{higgsmass}

We have discovered that the Higgs field in this scenario is not in the minimum of its potential but is hold towards a finite asymptotic value $\phi_{\rm asymp}$.  Thus, its mass today as well as the Yang-Mills fields' receive a contribution which is proportional to $\phi_{\rm asymp}$.  Moreover, in contrast to the SM, the Yang-Mills fields in this scenario do get a vacuum expectation value which approaches zero but never reaches it, so the Higgs mass and theirs also receive a contribution which is proportional to $f_0/a_0$.  Let's remember that, in the SM, the Higgs mass $m_\phi$ is proportional to the quartic coupling constant $\lambda$ whereas the Yang-Mills fields' $m_{A_\mu}$ is proportional to the SU(2) group coupling constant $\gamma$ through the following relations that involve the vacuum expectation value of the Higgs field $\phi_0$ \cite{Weinberg:1996kr,Kane:1987gb}:
\begin{eqnarray}
m_\phi &\sim& \sqrt{\lambda} \phi_0 \sim  \alpha \gamma \phi_0 \,, \\
m_{A_\mu} &\sim& \gamma \phi_0 \,.
\end{eqnarray}
In the present scenario:
\begin{eqnarray}
m_\phi &\sim& \gamma \sqrt{\alpha^2 \phi_{\rm asymp}^2 + \frac{1}{4} \frac{f_0^2}{a_0^2}} \,, \\
m_{A_\mu} &\sim& \gamma \sqrt{\phi_{\rm asymp}^2 + 4 \frac{f_0^2}{a_0^2}} \,.
\end{eqnarray}
where, according to Figs. \ref{higgsglobal} and \ref{vectordark}, $\phi_{\rm asymp} \approx 30 m_P$ and $f_0/a_0 \approx 10^{-2} m_P$.   On the other hand, from the dimensionless parameters in Eq. (\ref{variables}), we can obtain an expression for the Hubble parameter:
\begin{equation}
\frac{H}{m_P} = \sqrt{2} \frac{\gamma}{y l^2} \,,
\end{equation}
which can be used in conjunction with the measured value for the Hubble parameter today $H_0 \sim 10^{-61} m_P$ \cite{Aghanim:2018eyx} to obtain $\gamma \sim 10^{-75}$.  The present values of the Yang-Mills fields $A_\mu^a$ and the Higgs field $\phi$ are then $m_{A_\mu} \sim m_\phi \sim 10^{-74} m_P \sim 10^{-47}$ eV.  These are just figures since the actual values depend on the chosen initial conditions;  however, they give us a notion of how small $\gamma, m_\phi$ and $m_{A_\mu}$ must be. 


There are at least two ways of interpreting these results:  either 1. the extremely low value for $\gamma$ is a manifestation of the unsolved cosmological constant problem \cite{Weinberg:1988cp} so that the scenario discussed in this paper is just a `rephrasing' of such a difficult and fundamental problem in a more comprehensive particle physics context\footnote{It is very suggestive, however, that radiative corrections proportional to $\gamma$ do not enter in conflict with the measured value of $\Lambda$.}, or 2) this scenario is promising but requires some modification in order to have interactions beyond gravity between the $A_\mu^a$ and $\phi$ and the particles of the SM, and mass values for the $A_\mu^a$ and $\phi$ not so extremely low so that they can be scanned in future particle physics experiments.


\section{Conclusions} \label{conclusions}

The origin and nature of the dark energy have been elusive for physicists during two decades.  Most of the research has been concentrated either, in implementing a non-standard matter content (that violates the energy conditions) or in modifying the Einstein gravity.  M. Rinaldi's idea \cite{Rinaldi:2014yta} of studying the role, for dark energy, of the Higgs field and the Yang-Mills fields in a SM-like action with Einstein gravity is very clever since it avoids the usual avenues of research in the dark energy subject while feeding the analysis with well established and tested ideas from particle physics.  We have shown that the interaction between the Higgs field and the YangMills fields is fundamental for the success of this scenario, it being the key element that drives an asymptotic state that corresponds to the late accelerated expansion of the Universe.  Such an interaction is so important that, despite the fact that the Higgs field is a scalar and the Yang-Mills fields are in the cosmic triad configuration, the setup is highly anisotropic unless the gauge for the Higgs field is fixed so that the latter moves along a straight direction in the field space.  There can be no rotation around the axis of the Mexican hat potential and, therefore, this scenario differs completely from the ``spintessence'' one (see Ref. \cite{Boyle:2001du});  it is even different to the scenario in Ref. \cite{Rinaldi:2015iza}, where the SO(3) representation of the Higgs field is employed, since such a scenario is absolutely inconsistent with a homogeneous and isotropic universe \cite{newpaper}.  The usual stages in the evolution of the postinflationary universe are successfully reproduced, namely the radiation, matter, and dark energy dominated periods.  However, a new one, an early kination dominated period, is also obtained which would be very useful for the reheating process when modifying this model to implement primordial inflation \cite{Opferkuch:2019zbd}.
In addition, it is desirable that such a modification can explain the origin and nature of the dark matter and that implements interactions between the visible sector and the new one described in this paper, beyond the gravitational one, that allows us to test this scenario both from astronomical observations and in particle physics experiments.  We think this scenario represents a step towards a successful merging of cosmology and well-tested particle physics phenomenology.

\section*{Acknowledgements}
This work was supported by the following grants: Colciencias--DAAD -- 110278258747 RC-774-2017, Colciencias -- 123365843539 RC-FP44842-081-2014, Colciencias -- 110671250405 RC-FP44842-103-2016, DIEF de Ciencias - UIS - 2312, VCTI - UAN - 2017239, and Centro de Investigaciones - USTA - 1952392.  Y.R. wants to acknowledge Massimiliano Rinaldi for useful discussions related to this paper.

\appendix
\section{Complete list of eigenvalues and eigenvectors for the critical manifolds and critical points\footnote{In each item, the first number is the eigenvalue whereas the arrangement of numbers inside the curly brackets corresponds to the respective eigenvector.}} \label{appeigen}

{\it First critical manifold}: \\
$w = 0, z = 0, v = 0, r = \sqrt{1 - x^2 - y^2}, l = 0.$  
\begin{itemize}
\item 2, \{0, 0, 0, 0, 1, 0, 0\},
\item -1, \{0, 0, 0, 1, 0, 0, 0\},
\item 1, \{-$2y^2$, 2$ x y$, 0, 0, 
  0, 0, 1\},
\item 1, \{$x / \sqrt{1 - x^2 - y^2}$, 
   $y / \sqrt{1 - x^2 - y^2}$, 0, 0, 0, 1, 0\},
\item 1, \{0, 0, 1,
   0, 0, 0, 0\},
\item 0, \{-$\sqrt{1 - x^2 - y^2}/x$, 0, 0, 0, 0, 1, 0\},
\item 0, \{-$y/x$,
   1, 0, 0, 0, 0, 0\}.
\end{itemize}

{\it Second critical manifold}: \\
$y = 0, w = 0, z = 0, v = 0, r = \sqrt{1 - x^2}, l = 0.$  
\begin{itemize}
\item 2, \{0, 0, 0, 0, 1, 0, 0\},
\item -1, \{0, 0, 0, 1, 0, 0, 0\},
\item 1, \{0, 0, 0, 0, 0, 0, 1\},
\item 1, \{$x / \sqrt{1 - x^2}$, 0, 0, 0, 0, 1, 0\},
\item 1, \{0, 0, 1, 0, 0, 0, 0\},
\item 0, \{-$\sqrt{1 - x^2}/x$, 0, 0, 0, 0, 1, 0\},
\item 0, \{0, 1, 0, 0, 0, 0, 0\}.
\end{itemize}

{\it Third critical manifold}: \\
$y = 0, w = 0, z = 0, v = 0, r = \sqrt{1 - x^2}, l = 1/x.$  
\begin{itemize}
\item 0, \{-$x^2$, 0, 0, 0, 0, $x^3 / \sqrt{1 - x^2}$, 1\},
\item -1, \{0, 0, 0, 0, 0, 0, 1\},
\item -1, \{0, 0, 0, 1, 0, 0, 0\},
\item 1, \{-$2x^2$, 0, 0, 0, 0, -$2 x \sqrt{1 - x^2}$, 1\},
\item 2, \{0, 0, 0, 0, 1, 0, 0\},
\item 2, \{0, 0, 1, 0, 0, 0, 0\},
\item 2, \{0, 1, 0, 0, 0, 0, 0\}.
\end{itemize}

{\it Fourth critical manifold}: \\
$y = \sqrt{1 - x^2}, w = 0, z = 0, v = 0, r = 0, l = 0.$  
\begin{itemize}
\item 2, \{0, 0, 0, 0, 1, 0, 0\},
\item -1, \{0, 0, 0, 1, 0, 0, 0\},
\item 1, \{-2, 0, 0, 0, 0, 0, 1\},
\item 1, \{0, 0, 1, 0, 0, 0, 0\},
\item 1, \{$x / \sqrt{1 - x^2}$, 1, 0, 0, 0, 0, 0\},
\item 0, \{0, 0, 0, 0, 0, 1, 0\},
\item 0, \{-$\sqrt{1 - x^2}/x$, 1, 0, 0, 0, 0, 0\}.
\end{itemize}

{\it First critical point}: \\
$x = 0, y = 0, w = 1, z = 0, v = 0, r = 0, l = 0.$  
\begin{itemize}
\item -2, \{0, 0, 0, 1, 0, 0, 0\},
\item -1, \{0, 0, 0, 0, 0, 1, 0\},
\item -1, \{0, 0, 1, 0, 0, 0, 0\},
\item -1, \{0, 1, 0, 0, 0, 0, 0\},
\item -1, \{1, 0, 0, 0, 0, 0, 0\},
\item 1, \{1, 0, 0, 0, 0, 0, -2\},
\item 1, \{0, 0, 0, 0, -1, 0, 0\}.
\end{itemize}

{\it Second critical point}: \\
$x = 0, y = 0, w = 0, z = 1, v = 0, r = 0, l = 0.$  
\begin{itemize}
\item 3, \{0, 0, 0, 0, -1, 0, 0\},
\item 3, \{0, 0, 0, -1, 0, 0, 0\},
\item 2, \{0, 0, 1, 0, 0, 0, 0\},
\item 1, \{0, 0, 0, 0, 0, 0, 1\},
\item 1, \{0, 0, 0, 0, 0, 1, 0\},
\item 1, \{0, 1, 0, 0, 0, 0, 0\},
\item 1, \{1, 0, 0, 0, 0, 0, 0\}.
\end{itemize}

{\it Third critical point}: \\
$x = 0, y = 0, w = 0, z = 0, v = 1, r = 0, l = 0.$  
\begin{itemize}
\item -3, \{0, 0, 0, 0, 1, 0, 0\},
\item -3, \{0, 0, 0, 1, 0, 0, 0\},
\item -2, \{0, 0, 0, 0, 0, 1, 0\},
\item -2, \{0, 1, 0, 0, 0, 0, 0\},
\item -2, \{1, 0, 0, 0, 0, 0, 0\},
\item -1, \{0, 0, -1, 0, 0, 0, 0\},
\item 1, \{0, 0, 0, 0, 0, 0, 1\}.
\end{itemize}

{\it Fourth critical point}: \\
$x = 1, y = 0, w = 0, z = 0, v = 0, r = 0, l = 0.$  
\begin{itemize}
\item 2, \{0, 0, 0, 0, 1, 0, 0\},
\item -1, \{0, 0, 0, 1, 0, 0, 0\},
\item 1, \{0, 0, 0, 0, 0, 0, -1\},
\item 1, \{0, 0, -1, 0, 0, 0, 0\},
\item 1, \{-1, 0, 0, 0, 0, 0, 0\},
\item 0, \{0, 0, 0, 0, 0, -1, 0\},
\item 0, \{0, -1, 0, 0, 0, 0, 0\}.
\end{itemize}

{\it Fifth critical point}: \\
$x = 0, y = 0, w = 0, z = 0, v = 0, r = 0, l = 0.$  
\begin{itemize}
\item -$\frac{3}{2}$, \{0, 0, 0, 1, 0, 0, 0\},
\item $\frac{3}{2}$, \{0, 0, 0, 0, 1, 0, 0\},
\item 1, \{0, 0, 0, 0, 0, 0, 1\},
\item -$\frac{1}{2}$, \{0, 0, 0, 0, 0, 1, 0\},
\item -$\frac{1}{2}$, \{0, 1, 0, 0, 0, 0, 0\},
\item -$\frac{1}{2}$, \{1, 0, 0, 0, 0, 0, 0\},
\item $\frac{1}{2}$, \{0, 0, 1, 0, 0, 0, 0\}.
\end{itemize}

{\it Sixth critical point}: \\
$x = -1, y = 0, w = 0, z = 0, v = 0, r = 0, l = -1.$  
\begin{itemize}
\item 2, \{0, 0, 0, 0, 1, 0, 0\},
\item 2, \{0, 0, 1, 0, 0, 0, 0\},
\item 2, \{0, 1, 0, 0, 0, 0, 0\},
\item -1, \{0, 0, 0, 0, 0, 0, 1\},
\item -1, \{0, 0, 0, 1, 0, 0, 0\},
\item 1, \{2, 0, 0, 0, 0, 0, -1\},
\item 0, \{0, 0, 0, 0, 0, 1, 0\}.
\end{itemize}

{\it Seventh critical point}: \\
$x = 1, y = 0, w = 0, z = 0, v = 0, r = 0, l = 1.$  
\begin{itemize}
\item 2, \{0, 0, 0, 0, 1, 0, 0\},
\item 2, \{0, 0, 1, 0, 0, 0, 0\},
\item 2, \{0, 1, 0, 0, 0, 0, 0\},
\item -1, \{0, 0, 0, 0, 0, 0, 1\},
\item -1, \{0, 0, 0, 1, 0, 0, 0\},
\item 1, \{2, 0, 0, 0, 0, 0, -1\},
\item 0, \{0, 0, 0, 0, 0, 1, 0\}.
\end{itemize}

\section{The length of the radiation dominated period} \label{applengthrad}

The radiation fluid energy density evolves as $\rho_r = \rho_{r_i} (a_i/a)^4$, where the subindex $i$ represents some reference time.  The fact that this energy dominates during the radiation dominated period is represented by the Friedmann equation $H^2 = \rho_r/3m_P^2$.  Thus,
\begin{equation}
H = \sqrt{\frac{\rho_{r_i}}{3m_P^2}} \left( \frac{a_i}{a} \right)^2 \,,
\end{equation}
which can be readily integrated to give
\begin{equation}
a^2 = a_i^2 \left[ 1 + 2 \sqrt{\frac{\rho_{r_i}}{3m_P^2}} (t - t_i)  \right] \,. \label{aevolution}
\end{equation}
This equation, in turn, gives us the Hubble parameter as a function of time:
\begin{equation}
H(t) = \frac{\sqrt{\frac{\rho_{r_i}}{3m_P^2}}}{1 + 2 \sqrt{\frac{\rho_{r_i}}{3m_P^2}} (t - t_i)} \,. \label{Hevolution}
\end{equation}
Eqs. (\ref{aevolution}) and (\ref{Hevolution}) allow us to obtain the length in e-folds of the radiation dominated period (from the end of the kination dominated period to the time of matter-radiation equality):
\begin{equation}
\Delta N_{\rm rad} = - \ln \left( \frac{a_{\rm end}}{a_{\rm eq}} \right) \approx - \frac{1}{2} \ln \left( \frac{\sqrt{\frac{\rho_{r_{\rm eq}}}{3m_P^2}}}{H_{\rm end}} \right) \,,
\end{equation}
where the subscripts ``end'' and ``eq'' mean end of the kination dominated period and matter-radiation equality respectively.  Thus, $\Delta N_{\rm rad}$ can be written as
\begin{equation}
\Delta N_{\rm rad} \approx - \frac{1}{2} \ln \left( \sqrt{\Omega_{r_{\rm eq}}} \frac{H_{\rm eq}}{H_{\rm end}} \right) = - \frac{1}{2} \ln \left( \frac{H_{\rm eq}}{\sqrt{2} H_{\rm end}} \right) \,. \label{DNR}
\end{equation}
On the other hand, the matter fluid energy density evolves as $\rho_m = \rho_{m_i} (a_i/a)^3$, which leads to
\begin{equation}
H^2_{\rm eq} = \frac{2 \rho_{m_0}}{3 m_P^2} \left(\frac{a_0}{a_{\rm eq}}\right)^3 = 2 \Omega_{m_0} H_0^2 (1 + z_{\rm eq})^3 \,. \label{Heq}
\end{equation}
Introducing Eq. (\ref{Heq}) into Eq. (\ref{DNR}) gives us the final expression for $\Delta N_{\rm rad}$:
\begin{equation}
\Delta N_{\rm rad} \approx  - \frac{1}{2} \ln \left( \frac{\sqrt{\Omega_{m_0}} H_0 (1 + z_{\rm eq})^{3/2}}{H_{\rm end}} \right) \,.
\end{equation}

\bibliography{bibli.bib} 

\begin{thebibliography}{40}%
\makeatletter
\providecommand \@ifxundefined [1]{%
 \@ifx{#1\undefined}
}%
\providecommand \@ifnum [1]{%
 \ifnum #1\expandafter \@firstoftwo
 \else \expandafter \@secondoftwo
 \fi
}%
\providecommand \@ifx [1]{%
 \ifx #1\expandafter \@firstoftwo
 \else \expandafter \@secondoftwo
 \fi
}%
\providecommand \natexlab [1]{#1}%
\providecommand \enquote  [1]{``#1''}%
\providecommand \bibnamefont  [1]{#1}%
\providecommand \bibfnamefont [1]{#1}%
\providecommand \citenamefont [1]{#1}%
\providecommand \href@noop [0]{\@secondoftwo}%
\providecommand \href [0]{\begingroup \@sanitize@url \@href}%
\providecommand \@href[1]{\@@startlink{#1}\@@href}%
\providecommand \@@href[1]{\endgroup#1\@@endlink}%
\providecommand \@sanitize@url [0]{\catcode `\\12\catcode `\$12\catcode
  `\&12\catcode `\#12\catcode `\^12\catcode `\_12\catcode `\%12\relax}%
\providecommand \@@startlink[1]{}%
\providecommand \@@endlink[0]{}%
\providecommand \url  [0]{\begingroup\@sanitize@url \@url }%
\providecommand \@url [1]{\endgroup\@href {#1}{\urlprefix }}%
\providecommand \urlprefix  [0]{URL }%
\providecommand \Eprint [0]{\href }%
\providecommand \doibase [0]{http://dx.doi.org/}%
\providecommand \selectlanguage [0]{\@gobble}%
\providecommand \bibinfo  [0]{\@secondoftwo}%
\providecommand \bibfield  [0]{\@secondoftwo}%
\providecommand \translation [1]{[#1]}%
\providecommand \BibitemOpen [0]{}%
\providecommand \bibitemStop [0]{}%
\providecommand \bibitemNoStop [0]{.\EOS\space}%
\providecommand \EOS [0]{\spacefactor3000\relax}%
\providecommand \BibitemShut  [1]{\csname bibitem#1\endcsname}%
\let\auto@bib@innerbib\@empty
\bibitem [{\citenamefont {Riess}\ \emph {et~al.}(1998)\citenamefont {Riess}
  \emph {et~al.}}]{Riess:1998cb}%
  \BibitemOpen
  \bibfield  {author} {\bibinfo {author} {\bibfnamefont {A.~G.}\ \bibnamefont
  {Riess}} \emph {et~al.} (\bibinfo {collaboration} {Supernova Search Team}),\
  }\bibfield  {title} {\enquote {\bibinfo {title} {{Observational evidence from
  supernovae for an accelerating universe and a cosmological constant}},}\
  }\href {\doibase 10.1086/300499} {\bibfield  {journal} {\bibinfo  {journal}
  {Astron. J.}\ }\textbf {\bibinfo {volume} {116}},\ \bibinfo {pages}
  {1009--1038} (\bibinfo {year} {1998})},\ \Eprint
  {http://arxiv.org/abs/astro-ph/9805201} {arXiv:astro-ph/9805201} \BibitemShut
  {NoStop}%
\bibitem [{\citenamefont {Perlmutter}\ \emph {et~al.}(1999)\citenamefont
  {Perlmutter} \emph {et~al.}}]{Perlmutter:1998np}%
  \BibitemOpen
  \bibfield  {author} {\bibinfo {author} {\bibfnamefont {S.}~\bibnamefont
  {Perlmutter}} \emph {et~al.} (\bibinfo {collaboration} {Supernova Cosmology
  Project}),\ }\bibfield  {title} {\enquote {\bibinfo {title} {{Measurements of
  $\Omega$ and $\Lambda$ from 42 high redshift supernovae}},}\ }\href {\doibase
  10.1086/307221} {\bibfield  {journal} {\bibinfo  {journal} {Astrophys. J.}\
  }\textbf {\bibinfo {volume} {517}},\ \bibinfo {pages} {565--586} (\bibinfo
  {year} {1999})},\ \Eprint {http://arxiv.org/abs/astro-ph/9812133}
  {arXiv:astro-ph/9812133} \BibitemShut {NoStop}%
\bibitem [{\citenamefont {Huterer}\ and\ \citenamefont
  {Shafer}(2018)}]{Huterer:2017buf}%
  \BibitemOpen
  \bibfield  {author} {\bibinfo {author} {\bibfnamefont {D.}~\bibnamefont
  {Huterer}}\ and\ \bibinfo {author} {\bibfnamefont {D.~L.}\ \bibnamefont
  {Shafer}},\ }\bibfield  {title} {\enquote {\bibinfo {title} {{Dark energy two
  decades after: Observables, probes, consistency tests}},}\ }\href {\doibase
  10.1088/1361-6633/aa997e} {\bibfield  {journal} {\bibinfo  {journal} {Rept.
  Prog. Phys.}\ }\textbf {\bibinfo {volume} {81}},\ \bibinfo {pages} {016901}
  (\bibinfo {year} {2018})},\ \Eprint {http://arxiv.org/abs/1709.01091}
  {arXiv:1709.01091 [astro-ph.CO]} \BibitemShut {NoStop}%
\bibitem [{\citenamefont {Amendola}\ and\ \citenamefont
  {Tsujikawa}(2015)}]{Amendola:2015ksp}%
  \BibitemOpen
  \bibfield  {author} {\bibinfo {author} {\bibfnamefont {L.}~\bibnamefont
  {Amendola}}\ and\ \bibinfo {author} {\bibfnamefont {S.}~\bibnamefont
  {Tsujikawa}},\ }\href
  {http://www.cambridge.org/academic/subjects/physics/cosmology-relativity-and-gravitation/dark-energy-theory-and-observations?format=PB&isbn=9781107453982}
  {\emph {\bibinfo {title} {{Dark Energy}}}}\ (\bibinfo  {publisher} {Cambridge
  University Press},\ \bibinfo {year} {2015})\BibitemShut {NoStop}%
\bibitem [{\citenamefont {Aghanim}\ \emph {et~al.}(2018)\citenamefont {Aghanim}
  \emph {et~al.}}]{Aghanim:2018eyx}%
  \BibitemOpen
  \bibfield  {author} {\bibinfo {author} {\bibfnamefont {N.}~\bibnamefont
  {Aghanim}} \emph {et~al.} (\bibinfo {collaboration} {Planck}),\ }\bibfield
  {title} {\enquote {\bibinfo {title} {{Planck 2018 results. VI. Cosmological
  parameters}},}\ }\href@noop {} {\  (\bibinfo {year} {2018})},\ \Eprint
  {http://arxiv.org/abs/1807.06209} {arXiv:1807.06209 [astro-ph.CO]}
  \BibitemShut {NoStop}%
\bibitem [{\citenamefont {Weinberg}(1989)}]{Weinberg:1988cp}%
  \BibitemOpen
  \bibfield  {author} {\bibinfo {author} {\bibfnamefont {S.}~\bibnamefont
  {Weinberg}},\ }\bibfield  {title} {\enquote {\bibinfo {title} {{The
  Cosmological Constant Problem}},}\ }\href {\doibase 10.1103/RevModPhys.61.1}
  {\bibfield  {journal} {\bibinfo  {journal} {Rev. Mod. Phys.}\ }\textbf
  {\bibinfo {volume} {61}},\ \bibinfo {pages} {1--23} (\bibinfo {year}
  {1989})}\BibitemShut {NoStop}%
\bibitem [{\citenamefont {Collett}\ \emph {et~al.}(2018)\citenamefont {Collett}
  \emph {et~al.}}]{Collett:2018gpf}%
  \BibitemOpen
  \bibfield  {author} {\bibinfo {author} {\bibfnamefont {T.~E.}\ \bibnamefont
  {Collett}} \emph {et~al.},\ }\bibfield  {title} {\enquote {\bibinfo {title}
  {{A precise extragalactic test of General Relativity}},}\ }\href {\doibase
  10.1126/science.aao2469} {\bibfield  {journal} {\bibinfo  {journal}
  {Science}\ }\textbf {\bibinfo {volume} {360}},\ \bibinfo {pages} {1342}
  (\bibinfo {year} {2018})},\ \Eprint {http://arxiv.org/abs/1806.08300}
  {arXiv:1806.08300 [astro-ph.CO]} \BibitemShut {NoStop}%
\bibitem [{\citenamefont {Ezquiaga}\ and\ \citenamefont
  {Zumalacárregui}(2018)}]{Ezquiaga:2018btd}%
  \BibitemOpen
  \bibfield  {author} {\bibinfo {author} {\bibfnamefont {J.~M.}\ \bibnamefont
  {Ezquiaga}}\ and\ \bibinfo {author} {\bibfnamefont {M.}~\bibnamefont
  {Zumalacárregui}},\ }\bibfield  {title} {\enquote {\bibinfo {title} {{Dark
  Energy in light of Multi-Messenger Gravitational-Wave astronomy}},}\ }\href
  {\doibase 10.3389/fspas.2018.00044} {\bibfield  {journal} {\bibinfo
  {journal} {Front. Astron. Space Sci.}\ }\textbf {\bibinfo {volume} {5}},\
  \bibinfo {pages} {44} (\bibinfo {year} {2018})},\ \Eprint
  {http://arxiv.org/abs/1807.09241} {arXiv:1807.09241 [astro-ph.CO]}
  \BibitemShut {NoStop}%
\bibitem [{\citenamefont {He}\ \emph {et~al.}(2018)\citenamefont {He},
  \citenamefont {Guzzo}, \citenamefont {Li},\ and\ \citenamefont
  {Baugh}}]{He:2018oai}%
  \BibitemOpen
  \bibfield  {author} {\bibinfo {author} {\bibfnamefont {J.-h.}\ \bibnamefont
  {He}}, \bibinfo {author} {\bibfnamefont {L.}~\bibnamefont {Guzzo}}, \bibinfo
  {author} {\bibfnamefont {B.}~\bibnamefont {Li}}, \ and\ \bibinfo {author}
  {\bibfnamefont {C.~M.}\ \bibnamefont {Baugh}},\ }\bibfield  {title} {\enquote
  {\bibinfo {title} {{No evidence for modifications of gravity from galaxy
  motions on cosmological scales}},}\ }\href {\doibase
  10.1038/s41550-018-0573-2} {\bibfield  {journal} {\bibinfo  {journal} {Nat.
  Astron.}\ }\textbf {\bibinfo {volume} {2}},\ \bibinfo {pages} {967--972}
  (\bibinfo {year} {2018})},\ \Eprint {http://arxiv.org/abs/1809.09019}
  {arXiv:1809.09019 [astro-ph.CO]} \BibitemShut {NoStop}%
\bibitem [{\citenamefont {Weinberg}(1996)}]{Weinberg:1996kr}%
  \BibitemOpen
  \bibfield  {author} {\bibinfo {author} {\bibfnamefont {S.}~\bibnamefont
  {Weinberg}},\ }\href
  {http://www.cambridge.org/core/books/quantum-theory-of-fields/0E0C89894938BE38EE0BCCDB1BC857E5}
  {\emph {\bibinfo {title} {{The quantum theory of fields. Vol. 2: Modern
  applications}}}}\ (\bibinfo  {publisher} {Cambridge University Press},\
  \bibinfo {year} {1996})\BibitemShut {NoStop}%
\bibitem [{\citenamefont {Kane}(2017)}]{Kane:1987gb}%
  \BibitemOpen
  \bibfield  {author} {\bibinfo {author} {\bibfnamefont {G.~L.}\ \bibnamefont
  {Kane}},\ }\href
  {http://www.cambridge.org/academic/subjects/physics/particle-physics-and-nuclear-physics/modern-elementary-particle-physics-explaining-and-extending-standard-model-2nd-edition?format=AR&isbn=9781316730805}
  {\emph {\bibinfo {title} {{Modern Elementary Particle Physics}}}}\ (\bibinfo
  {publisher} {Cambridge University Press},\ \bibinfo {year}
  {2017})\BibitemShut {NoStop}%
\bibitem [{\citenamefont {Moniz}\ \emph {et~al.}(1993)\citenamefont {Moniz},
  \citenamefont {Mourao},\ and\ \citenamefont {Sa}}]{Moniz:1991kx}%
  \BibitemOpen
  \bibfield  {author} {\bibinfo {author} {\bibfnamefont {P.~V.}\ \bibnamefont
  {Moniz}}, \bibinfo {author} {\bibfnamefont {J.~M.}\ \bibnamefont {Mourao}}, \
  and\ \bibinfo {author} {\bibfnamefont {P.~M.}\ \bibnamefont {Sa}},\
  }\bibfield  {title} {\enquote {\bibinfo {title} {{The Dynamics of a flat
  Friedmann-Robertson-Walker inflationary model in the presence of gauge
  fields}},}\ }\href {\doibase 10.1088/0264-9381/10/3/012} {\bibfield
  {journal} {\bibinfo  {journal} {Class. Quantum Grav.}\ }\textbf {\bibinfo
  {volume} {10}},\ \bibinfo {pages} {517--534} (\bibinfo {year}
  {1993})}\BibitemShut {NoStop}%
\bibitem [{\citenamefont {Ochs}\ and\ \citenamefont
  {Sorg}(1996)}]{Ochs:1996yr}%
  \BibitemOpen
  \bibfield  {author} {\bibinfo {author} {\bibfnamefont {U.}~\bibnamefont
  {Ochs}}\ and\ \bibinfo {author} {\bibfnamefont {M.}~\bibnamefont {Sorg}},\
  }\bibfield  {title} {\enquote {\bibinfo {title} {{Cosmological solutions for
  the coupled Einstein-Yang-Mills-Higgs equations}},}\ }\href {\doibase
  10.1007/BF02107381} {\bibfield  {journal} {\bibinfo  {journal} {Gen. Rel.
  Grav.}\ }\textbf {\bibinfo {volume} {28}},\ \bibinfo {pages} {1177--1219}
  (\bibinfo {year} {1996})}\BibitemShut {NoStop}%
\bibitem [{\citenamefont {Rinaldi}(2015{\natexlab{a}})}]{Rinaldi:2014yta}%
  \BibitemOpen
  \bibfield  {author} {\bibinfo {author} {\bibfnamefont {M.}~\bibnamefont
  {Rinaldi}},\ }\bibfield  {title} {\enquote {\bibinfo {title} {{Higgs Dark
  Energy}},}\ }\href {\doibase 10.1088/0264-9381/32/4/045002} {\bibfield
  {journal} {\bibinfo  {journal} {Class. Quantum Grav.}\ }\textbf {\bibinfo
  {volume} {32}},\ \bibinfo {pages} {045002} (\bibinfo {year}
  {2015}{\natexlab{a}})},\ \Eprint {http://arxiv.org/abs/1404.0532}
  {arXiv:1404.0532 [astro-ph.CO]} \BibitemShut {NoStop}%
\bibitem [{\citenamefont {Aad}\ \emph {et~al.}(2012)\citenamefont {Aad} \emph
  {et~al.}}]{Aad:2012tfa}%
  \BibitemOpen
  \bibfield  {author} {\bibinfo {author} {\bibfnamefont {G.}~\bibnamefont
  {Aad}} \emph {et~al.} (\bibinfo {collaboration} {ATLAS}),\ }\bibfield
  {title} {\enquote {\bibinfo {title} {{Observation of a new particle in the
  search for the Standard Model Higgs boson with the ATLAS detector at the
  LHC}},}\ }\href {\doibase 10.1016/j.physletb.2012.08.020} {\bibfield
  {journal} {\bibinfo  {journal} {Phys. Lett. B}\ }\textbf {\bibinfo {volume}
  {716}},\ \bibinfo {pages} {1--29} (\bibinfo {year} {2012})},\ \Eprint
  {http://arxiv.org/abs/1207.7214} {arXiv:1207.7214 [hep-ex]} \BibitemShut
  {NoStop}%
\bibitem [{\citenamefont {Chatrchyan}\ \emph {et~al.}(2012)\citenamefont
  {Chatrchyan} \emph {et~al.}}]{Chatrchyan:2012xdj}%
  \BibitemOpen
  \bibfield  {author} {\bibinfo {author} {\bibfnamefont {S.}~\bibnamefont
  {Chatrchyan}} \emph {et~al.} (\bibinfo {collaboration} {CMS}),\ }\bibfield
  {title} {\enquote {\bibinfo {title} {{Observation of a new boson at a mass of
  125 GeV with the CMS experiment at the LHC}},}\ }\href {\doibase
  10.1016/j.physletb.2012.08.021} {\bibfield  {journal} {\bibinfo  {journal}
  {Phys. Lett. B}\ }\textbf {\bibinfo {volume} {716}},\ \bibinfo {pages}
  {30--61} (\bibinfo {year} {2012})},\ \Eprint {http://arxiv.org/abs/1207.7235}
  {arXiv:1207.7235 [hep-ex]} \BibitemShut {NoStop}%
\bibitem [{\citenamefont {Boyle}\ \emph {et~al.}(2002)\citenamefont {Boyle},
  \citenamefont {Caldwell},\ and\ \citenamefont {Kamionkowski}}]{Boyle:2001du}%
  \BibitemOpen
  \bibfield  {author} {\bibinfo {author} {\bibfnamefont {L.~A.}\ \bibnamefont
  {Boyle}}, \bibinfo {author} {\bibfnamefont {R.~R.}\ \bibnamefont {Caldwell}},
  \ and\ \bibinfo {author} {\bibfnamefont {M.}~\bibnamefont {Kamionkowski}},\
  }\bibfield  {title} {\enquote {\bibinfo {title} {{Spintessence! New models
  for dark matter and dark energy}},}\ }\href {\doibase
  10.1016/S0370-2693(02)02590-X} {\bibfield  {journal} {\bibinfo  {journal}
  {Phys. Lett. B}\ }\textbf {\bibinfo {volume} {545}},\ \bibinfo {pages}
  {17--22} (\bibinfo {year} {2002})},\ \Eprint
  {http://arxiv.org/abs/astro-ph/0105318} {arXiv:astro-ph/0105318} \BibitemShut
  {NoStop}%
\bibitem [{\citenamefont {Rinaldi}(2015{\natexlab{b}})}]{Rinaldi:2015iza}%
  \BibitemOpen
  \bibfield  {author} {\bibinfo {author} {\bibfnamefont {M.}~\bibnamefont
  {Rinaldi}},\ }\bibfield  {title} {\enquote {\bibinfo {title} {{Dark energy as
  a fixed point of the Einstein Yang-Mills Higgs Equations}},}\ }\href
  {\doibase 10.1088/1475-7516/2015/10/023} {\bibfield  {journal} {\bibinfo
  {journal} {JCAP}\ }\textbf {\bibinfo {volume} {1510}},\ \bibinfo {pages}
  {023} (\bibinfo {year} {2015}{\natexlab{b}})},\ \Eprint
  {http://arxiv.org/abs/1508.04576} {arXiv:1508.04576 [gr-qc]} \BibitemShut
  {NoStop}%
\bibitem [{\citenamefont {Orjuela-Quintana}\ \emph {et~al.}(2019, Work in
  progress)\citenamefont {Orjuela-Quintana}, \citenamefont {Álvarez},
  \citenamefont {Valenzuela-Toledo},\ and\ \citenamefont
  {Rodríguez}}]{newpaper}%
  \BibitemOpen
  \bibfield  {author} {\bibinfo {author} {\bibfnamefont {J.~B.}\ \bibnamefont
  {Orjuela-Quintana}}, \bibinfo {author} {\bibfnamefont {M.}~\bibnamefont
  {Álvarez}}, \bibinfo {author} {\bibfnamefont {C.~A.}\ \bibnamefont
  {Valenzuela-Toledo}}, \ and\ \bibinfo {author} {\bibfnamefont
  {Y.}~\bibnamefont {Rodríguez}},\ }\href@noop {} {\  (\bibinfo {year} {2019,
  Work in progress})}\BibitemShut {NoStop}%
\bibitem [{\citenamefont {Armendariz-Picon}(2004)}]{ArmendarizPicon:2004pm}%
  \BibitemOpen
  \bibfield  {author} {\bibinfo {author} {\bibfnamefont {C.}~\bibnamefont
  {Armendariz-Picon}},\ }\bibfield  {title} {\enquote {\bibinfo {title} {{Could
  dark energy be vector-like?}}}\ }\href {\doibase
  10.1088/1475-7516/2004/07/007} {\bibfield  {journal} {\bibinfo  {journal}
  {JCAP}\ }\textbf {\bibinfo {volume} {0407}},\ \bibinfo {pages} {007}
  (\bibinfo {year} {2004})},\ \Eprint {http://arxiv.org/abs/astro-ph/0405267}
  {arXiv:astro-ph/0405267} \BibitemShut {NoStop}%
\bibitem [{\citenamefont {Mehrabi}\ \emph {et~al.}(2017)\citenamefont
  {Mehrabi}, \citenamefont {Maleknejad},\ and\ \citenamefont
  {Kamali}}]{Mehrabi:2017xga}%
  \BibitemOpen
  \bibfield  {author} {\bibinfo {author} {\bibfnamefont {A.}~\bibnamefont
  {Mehrabi}}, \bibinfo {author} {\bibfnamefont {A.}~\bibnamefont {Maleknejad}},
  \ and\ \bibinfo {author} {\bibfnamefont {V.}~\bibnamefont {Kamali}},\
  }\bibfield  {title} {\enquote {\bibinfo {title} {{Gaugessence: a dark energy
  model with early time radiation-like equation of state}},}\ }\href {\doibase
  10.1007/s10509-017-3033-z} {\bibfield  {journal} {\bibinfo  {journal}
  {Astrophys. Space Sci.}\ }\textbf {\bibinfo {volume} {362}},\ \bibinfo
  {pages} {53} (\bibinfo {year} {2017})},\ \Eprint
  {http://arxiv.org/abs/1510.00838} {arXiv:1510.00838 [astro-ph.CO]}
  \BibitemShut {NoStop}%
\bibitem [{\citenamefont {Rodríguez}\ and\ \citenamefont
  {Navarro}(2018)}]{Rodriguez:2017wkg}%
  \BibitemOpen
  \bibfield  {author} {\bibinfo {author} {\bibfnamefont {Y.}~\bibnamefont
  {Rodríguez}}\ and\ \bibinfo {author} {\bibfnamefont {A.~A.}\ \bibnamefont
  {Navarro}},\ }\bibfield  {title} {\enquote {\bibinfo {title} {{Non-Abelian
  $S$-term dark energy and inflation}},}\ }\href {\doibase
  10.1016/j.dark.2018.01.003} {\bibfield  {journal} {\bibinfo  {journal} {Phys.
  Dark Univ.}\ }\textbf {\bibinfo {volume} {19}},\ \bibinfo {pages} {129--136}
  (\bibinfo {year} {2018})},\ \Eprint {http://arxiv.org/abs/1711.01935}
  {arXiv:1711.01935 [gr-qc]} \BibitemShut {NoStop}%
\bibitem [{\citenamefont {Endlich}\ \emph {et~al.}(2013)\citenamefont
  {Endlich}, \citenamefont {Nicolis},\ and\ \citenamefont
  {Wang}}]{Endlich:2012pz}%
  \BibitemOpen
  \bibfield  {author} {\bibinfo {author} {\bibfnamefont {S.}~\bibnamefont
  {Endlich}}, \bibinfo {author} {\bibfnamefont {A.}~\bibnamefont {Nicolis}}, \
  and\ \bibinfo {author} {\bibfnamefont {J.}~\bibnamefont {Wang}},\ }\bibfield
  {title} {\enquote {\bibinfo {title} {{Solid Inflation}},}\ }\href {\doibase
  10.1088/1475-7516/2013/10/011} {\bibfield  {journal} {\bibinfo  {journal}
  {JCAP}\ }\textbf {\bibinfo {volume} {1310}},\ \bibinfo {pages} {011}
  (\bibinfo {year} {2013})},\ \Eprint {http://arxiv.org/abs/1210.0569}
  {arXiv:1210.0569 [hep-th]} \BibitemShut {NoStop}%
\bibitem [{\citenamefont {Bartolo}\ \emph {et~al.}(2013)\citenamefont
  {Bartolo}, \citenamefont {Matarrese}, \citenamefont {Peloso},\ and\
  \citenamefont {Ricciardone}}]{Bartolo:2013msa}%
  \BibitemOpen
  \bibfield  {author} {\bibinfo {author} {\bibfnamefont {N.}~\bibnamefont
  {Bartolo}}, \bibinfo {author} {\bibfnamefont {S.}~\bibnamefont {Matarrese}},
  \bibinfo {author} {\bibfnamefont {M.}~\bibnamefont {Peloso}}, \ and\ \bibinfo
  {author} {\bibfnamefont {A.}~\bibnamefont {Ricciardone}},\ }\bibfield
  {title} {\enquote {\bibinfo {title} {{Anisotropy in solid inflation}},}\
  }\href {\doibase 10.1088/1475-7516/2013/08/022} {\bibfield  {journal}
  {\bibinfo  {journal} {JCAP}\ }\textbf {\bibinfo {volume} {1308}},\ \bibinfo
  {pages} {022} (\bibinfo {year} {2013})},\ \Eprint
  {http://arxiv.org/abs/1306.4160} {arXiv:1306.4160 [astro-ph.CO]} \BibitemShut
  {NoStop}%
\bibitem [{\citenamefont {Rodríguez}\ \emph {et~al.}(2015)\citenamefont
  {Rodríguez}, \citenamefont {Gómez},\ and\ \citenamefont
  {Nieto}}]{Rodriguez:2015xra}%
  \BibitemOpen
  \bibfield  {author} {\bibinfo {author} {\bibfnamefont {Y.}~\bibnamefont
  {Rodríguez}}, \bibinfo {author} {\bibfnamefont {L.~G.}\ \bibnamefont
  {Gómez}}, \ and\ \bibinfo {author} {\bibfnamefont {C.~M.}\ \bibnamefont
  {Nieto}},\ }\bibfield  {title} {\enquote {\bibinfo {title} {{Towards a proof
  of the equivalence between FRW background expansion and statistical
  isotropy}},}\ }in\ \href@noop {} {\emph {\bibinfo {booktitle} {{Proceedings,
  2nd Argentinian-Brazilian Meeting on Gravitation, Relativistic Astrophysics
  and Cosmology (GRACo II): Buenos Aires, Argentina, April 22-25, 2014}}}}\
  (\bibinfo {year} {2015})\ pp.\ \bibinfo {pages} {139--144},\ \Eprint
  {http://arxiv.org/abs/1502.07292} {arXiv:1502.07292 [astro-ph.CO]}
  \BibitemShut {NoStop}%
\bibitem [{\citenamefont {Golovnev}\ \emph {et~al.}(2008)\citenamefont
  {Golovnev}, \citenamefont {Mukhanov},\ and\ \citenamefont
  {Vanchurin}}]{Golovnev:2008cf}%
  \BibitemOpen
  \bibfield  {author} {\bibinfo {author} {\bibfnamefont {A.}~\bibnamefont
  {Golovnev}}, \bibinfo {author} {\bibfnamefont {V.}~\bibnamefont {Mukhanov}},
  \ and\ \bibinfo {author} {\bibfnamefont {V.}~\bibnamefont {Vanchurin}},\
  }\bibfield  {title} {\enquote {\bibinfo {title} {{Vector Inflation}},}\
  }\href {\doibase 10.1088/1475-7516/2008/06/009} {\bibfield  {journal}
  {\bibinfo  {journal} {JCAP}\ }\textbf {\bibinfo {volume} {0806}},\ \bibinfo
  {pages} {009} (\bibinfo {year} {2008})},\ \Eprint
  {http://arxiv.org/abs/0802.2068} {arXiv:0802.2068 [astro-ph]} \BibitemShut
  {NoStop}%
\bibitem [{\citenamefont {Maleknejad}\ and\ \citenamefont
  {Sheikh-Jabbari}(2013)}]{Maleknejad:2011jw}%
  \BibitemOpen
  \bibfield  {author} {\bibinfo {author} {\bibfnamefont {A.}~\bibnamefont
  {Maleknejad}}\ and\ \bibinfo {author} {\bibfnamefont {M.~M.}\ \bibnamefont
  {Sheikh-Jabbari}},\ }\bibfield  {title} {\enquote {\bibinfo {title}
  {{Gauge-flation: Inflation From Non-Abelian Gauge Fields}},}\ }\href
  {\doibase 10.1016/j.physletb.2013.05.001} {\bibfield  {journal} {\bibinfo
  {journal} {Phys. Lett. B}\ }\textbf {\bibinfo {volume} {723}},\ \bibinfo
  {pages} {224--228} (\bibinfo {year} {2013})},\ \Eprint
  {http://arxiv.org/abs/1102.1513} {arXiv:1102.1513 [hep-ph]} \BibitemShut
  {NoStop}%
\bibitem [{\citenamefont {Nieto}\ and\ \citenamefont
  {Rodríguez}(2016)}]{Nieto:2016gnp}%
  \BibitemOpen
  \bibfield  {author} {\bibinfo {author} {\bibfnamefont {C.~M.}\ \bibnamefont
  {Nieto}}\ and\ \bibinfo {author} {\bibfnamefont {Y.}~\bibnamefont
  {Rodríguez}},\ }\bibfield  {title} {\enquote {\bibinfo {title} {{Massive
  Gauge-flation}},}\ }\href {\doibase 10.1142/S0217732316400058} {\bibfield
  {journal} {\bibinfo  {journal} {Mod. Phys. Lett. A}\ }\textbf {\bibinfo
  {volume} {31}},\ \bibinfo {pages} {1640005} (\bibinfo {year} {2016})},\
  \Eprint {http://arxiv.org/abs/1602.07197} {arXiv:1602.07197 [gr-qc]}
  \BibitemShut {NoStop}%
\bibitem [{\citenamefont {Wagstaff}\ and\ \citenamefont
  {Dimopoulos}(2011)}]{Dimopoulos:2010xq}%
  \BibitemOpen
  \bibfield  {author} {\bibinfo {author} {\bibfnamefont {J.~M.}\ \bibnamefont
  {Wagstaff}}\ and\ \bibinfo {author} {\bibfnamefont {K.}~\bibnamefont
  {Dimopoulos}},\ }\bibfield  {title} {\enquote {\bibinfo {title} {{Particle
  Production of Vector Fields: Scale Invariance is Attractive}},}\ }\href
  {\doibase 10.1103/PhysRevD.83.023523} {\bibfield  {journal} {\bibinfo
  {journal} {Phys. Rev.}\ }\textbf {\bibinfo {volume} {D83}},\ \bibinfo {pages}
  {023523} (\bibinfo {year} {2011})},\ \Eprint {http://arxiv.org/abs/1011.2517}
  {arXiv:1011.2517 [hep-ph]} \BibitemShut {NoStop}%
\bibitem [{\citenamefont {Lyth}(2017)}]{Lyth:2017hyk}%
  \BibitemOpen
  \bibfield  {author} {\bibinfo {author} {\bibfnamefont {D.~H.}\ \bibnamefont
  {Lyth}},\ }\href
  {https://www.crcpress.com/Cosmology-for-Physicists/Lyth/p/book/9781498755313}
  {\emph {\bibinfo {title} {{Cosmology for physicists}}}}\ (\bibinfo
  {publisher} {CRC Press},\ \bibinfo {address} {Boca Raton},\ \bibinfo {year}
  {2017})\BibitemShut {NoStop}%
\bibitem [{\citenamefont {Akrami}\ \emph {et~al.}(2018)\citenamefont {Akrami}
  \emph {et~al.}}]{Akrami:2018odb}%
  \BibitemOpen
  \bibfield  {author} {\bibinfo {author} {\bibfnamefont {Y.}~\bibnamefont
  {Akrami}} \emph {et~al.} (\bibinfo {collaboration} {Planck}),\ }\bibfield
  {title} {\enquote {\bibinfo {title} {{Planck 2018 results. X. Constraints on
  inflation}},}\ }\href@noop {} {\  (\bibinfo {year} {2018})},\ \Eprint
  {http://arxiv.org/abs/1807.06211} {arXiv:1807.06211 [astro-ph.CO]}
  \BibitemShut {NoStop}%
\bibitem [{\citenamefont {Opferkuch}\ \emph {et~al.}(2019)\citenamefont
  {Opferkuch}, \citenamefont {Schwaller},\ and\ \citenamefont
  {Stefanek}}]{Opferkuch:2019zbd}%
  \BibitemOpen
  \bibfield  {author} {\bibinfo {author} {\bibfnamefont {T.}~\bibnamefont
  {Opferkuch}}, \bibinfo {author} {\bibfnamefont {P.}~\bibnamefont
  {Schwaller}}, \ and\ \bibinfo {author} {\bibfnamefont {B.~A.}\ \bibnamefont
  {Stefanek}},\ }\bibfield  {title} {\enquote {\bibinfo {title} {{Ricci
  Reheating}},}\ }\href {\doibase 10.1088/1475-7516/2019/07/016} {\bibfield
  {journal} {\bibinfo  {journal} {JCAP}\ }\textbf {\bibinfo {volume} {1907}},\
  \bibinfo {pages} {016} (\bibinfo {year} {2019})},\ \Eprint
  {http://arxiv.org/abs/1905.06823} {arXiv:1905.06823 [gr-qc]} \BibitemShut
  {NoStop}%
\bibitem [{\citenamefont {Rajvanshi}\ and\ \citenamefont
  {Bagla}(2019)}]{Rajvanshi:2019wmw}%
  \BibitemOpen
  \bibfield  {author} {\bibinfo {author} {\bibfnamefont {M.~P.}\ \bibnamefont
  {Rajvanshi}}\ and\ \bibinfo {author} {\bibfnamefont {J.~S.}\ \bibnamefont
  {Bagla}},\ }\bibfield  {title} {\enquote {\bibinfo {title} {{Reconstruction
  of Dynamical Dark Energy Potentials: Quintessence, Tachyon and interacting
  models}},}\ }\href@noop {} {\  (\bibinfo {year} {2019})},\ \Eprint
  {http://arxiv.org/abs/1905.01103} {arXiv:1905.01103 [astro-ph.CO]}
  \BibitemShut {NoStop}%
\bibitem [{\citenamefont {Chevallier}\ and\ \citenamefont
  {Polarski}(2001)}]{Chevallier:2000qy}%
  \BibitemOpen
  \bibfield  {author} {\bibinfo {author} {\bibfnamefont {M.}~\bibnamefont
  {Chevallier}}\ and\ \bibinfo {author} {\bibfnamefont {D.}~\bibnamefont
  {Polarski}},\ }\bibfield  {title} {\enquote {\bibinfo {title} {{Accelerating
  universes with scaling dark matter}},}\ }\href {\doibase
  10.1142/S0218271801000822} {\bibfield  {journal} {\bibinfo  {journal} {Int.
  J. Mod. Phys.}\ }\textbf {\bibinfo {volume} {D10}},\ \bibinfo {pages}
  {213--224} (\bibinfo {year} {2001})},\ \Eprint
  {http://arxiv.org/abs/gr-qc/0009008} {arXiv:gr-qc/0009008 [gr-qc]}
  \BibitemShut {NoStop}%
\bibitem [{\citenamefont {Linder}(2003)}]{Linder:2002et}%
  \BibitemOpen
  \bibfield  {author} {\bibinfo {author} {\bibfnamefont {E.~V.}\ \bibnamefont
  {Linder}},\ }\bibfield  {title} {\enquote {\bibinfo {title} {{Exploring the
  expansion history of the universe}},}\ }\href {\doibase
  10.1103/PhysRevLett.90.091301} {\bibfield  {journal} {\bibinfo  {journal}
  {Phys. Rev. Lett.}\ }\textbf {\bibinfo {volume} {90}},\ \bibinfo {pages}
  {091301} (\bibinfo {year} {2003})},\ \Eprint
  {http://arxiv.org/abs/astro-ph/0208512} {arXiv:astro-ph/0208512 [astro-ph]}
  \BibitemShut {NoStop}%
\bibitem [{\citenamefont {Kase}\ and\ \citenamefont
  {Tsujikawa}(2019)}]{Kase:2018aps}%
  \BibitemOpen
  \bibfield  {author} {\bibinfo {author} {\bibfnamefont {R.}~\bibnamefont
  {Kase}}\ and\ \bibinfo {author} {\bibfnamefont {S.}~\bibnamefont
  {Tsujikawa}},\ }\bibfield  {title} {\enquote {\bibinfo {title} {{Dark energy
  in Horndeski theories after GW170817: A review}},}\ }\href {\doibase
  10.1142/S0218271819420057} {\bibfield  {journal} {\bibinfo  {journal} {Int.
  J. Mod. Phys.}\ }\textbf {\bibinfo {volume} {D28}},\ \bibinfo {pages}
  {1942005} (\bibinfo {year} {2019})},\ \Eprint
  {http://arxiv.org/abs/1809.08735} {arXiv:1809.08735 [gr-qc]} \BibitemShut
  {NoStop}%
\bibitem [{\citenamefont {Bartolo}\ \emph {et~al.}(2018)\citenamefont {Bartolo}
  \emph {et~al.}}]{Bartolo:2017sbu}%
  \BibitemOpen
  \bibfield  {author} {\bibinfo {author} {\bibfnamefont {N.}~\bibnamefont
  {Bartolo}} \emph {et~al.},\ }\bibfield  {title} {\enquote {\bibinfo {title}
  {{Detecting higher spin fields through statistical anisotropy in the CMB and
  galaxy power spectra}},}\ }\href {\doibase 10.1103/PhysRevD.97.023503}
  {\bibfield  {journal} {\bibinfo  {journal} {Phys. Rev.}\ }\textbf {\bibinfo
  {volume} {D97}},\ \bibinfo {pages} {023503} (\bibinfo {year} {2018})},\
  \Eprint {http://arxiv.org/abs/1709.05695} {arXiv:1709.05695 [astro-ph.CO]}
  \BibitemShut {NoStop}%
\bibitem [{\citenamefont {Bonvin}\ \emph {et~al.}(2018)\citenamefont {Bonvin}
  \emph {et~al.}}]{Bonvin:2017req}%
  \BibitemOpen
  \bibfield  {author} {\bibinfo {author} {\bibfnamefont {C.}~\bibnamefont
  {Bonvin}} \emph {et~al.},\ }\bibfield  {title} {\enquote {\bibinfo {title}
  {{Redshift-space distortions from vector perturbations}},}\ }\href {\doibase
  10.1088/1475-7516/2018/02/028} {\bibfield  {journal} {\bibinfo  {journal}
  {JCAP}\ }\textbf {\bibinfo {volume} {1802}},\ \bibinfo {pages} {028}
  (\bibinfo {year} {2018})},\ \Eprint {http://arxiv.org/abs/1712.00052}
  {arXiv:1712.00052 [astro-ph.CO]} \BibitemShut {NoStop}%
\bibitem [{\citenamefont {Moradinezhad~Dizgah}\ \emph
  {et~al.}(2018)\citenamefont {Moradinezhad~Dizgah}, \citenamefont
  {Franciolini}, \citenamefont {Kehagias},\ and\ \citenamefont
  {Riotto}}]{MoradinezhadDizgah:2018pfo}%
  \BibitemOpen
  \bibfield  {author} {\bibinfo {author} {\bibfnamefont {A.}~\bibnamefont
  {Moradinezhad~Dizgah}}, \bibinfo {author} {\bibfnamefont {G.}~\bibnamefont
  {Franciolini}}, \bibinfo {author} {\bibfnamefont {A.}~\bibnamefont
  {Kehagias}}, \ and\ \bibinfo {author} {\bibfnamefont {A.}~\bibnamefont
  {Riotto}},\ }\bibfield  {title} {\enquote {\bibinfo {title} {{Constraints on
  long-lived, higher-spin particles from galaxy bispectrum}},}\ }\href
  {\doibase 10.1103/PhysRevD.98.063520} {\bibfield  {journal} {\bibinfo
  {journal} {Phys. Rev.}\ }\textbf {\bibinfo {volume} {D98}},\ \bibinfo {pages}
  {063520} (\bibinfo {year} {2018})},\ \Eprint
  {http://arxiv.org/abs/1805.10247} {arXiv:1805.10247 [astro-ph.CO]}
  \BibitemShut {NoStop}%
\bibitem [{\citenamefont {Tansella}\ \emph {et~al.}(2018)\citenamefont
  {Tansella} \emph {et~al.}}]{Tansella:2018hdm}%
  \BibitemOpen
  \bibfield  {author} {\bibinfo {author} {\bibfnamefont {V.}~\bibnamefont
  {Tansella}} \emph {et~al.},\ }\bibfield  {title} {\enquote {\bibinfo {title}
  {{Redshift-space distortions from vector perturbations II: Anisotropic
  signal}},}\ }\href {\doibase 10.1103/PhysRevD.98.103515} {\bibfield
  {journal} {\bibinfo  {journal} {Phys. Rev.}\ }\textbf {\bibinfo {volume}
  {D98}},\ \bibinfo {pages} {103515} (\bibinfo {year} {2018})},\ \Eprint
  {http://arxiv.org/abs/1807.00731} {arXiv:1807.00731 [astro-ph.CO]}
  \BibitemShut {NoStop}%
\end{thebibliography}%

\end{document}